\documentclass[a4paper,11pt]{article} \pdfoutput=1 \usepackage{jheppub}
\usepackage{bm} \usepackage{bbold} \usepackage{multirow} \usepackage{diagbox}
\usepackage{lineno} \usepackage[normalem]{ulem} %\linenumbers

\usepackage[T1]{fontenc} \usepackage[pdftex]{color}

\allowdisplaybreaks

\title{\boldmath QCD Static Force in Gradient Flow } \preprint{TUM-EFT 152/21}

\author[a,b,c]{Nora~Brambilla,} \author[a,d]{Hee~Sok~Chung,}
\author[a]{Antonio~Vairo,} \author[a]{and Xiang-Peng~Wang}

\affiliation[a]{Physik Department, Technische Universit\"at M\"unchen,\\
James-Franck-Strasse 1, 85748 Garching, Germany} \affiliation[b]{Institute for
Advanced Study, Technische Universit\"at M\"unchen,\\ Lichtenbergstrasse 2 a,
85748 Garching, Germany} \affiliation[c]{Munich Data Science Institute,
Technische Universit\"at M\"unchen, \\ Walther-von-Dyck-Strasse 10, 85748
Garching, Germany} \affiliation[d]{Excellence Cluster ORIGINS, Boltzmannstrasse
2, 85748 Garching, Germany}

\emailAdd{nora.brambilla@tum.de} \emailAdd{heesok.chung@tum.de}
\emailAdd{antonio.vairo@tum.de} \emailAdd{xiangpeng.wang@tum.de}

\abstract{ We compute the QCD static force and potential using gradient
flow at next-to-leading order in the strong coupling.  The static force is the
spatial derivative of the static potential: it encodes
the QCD interaction at both short and long distances.  While on
the one side the static force has the advantage of being free of the
$O(\Lambda_{\rm QCD})$ renormalon affecting the static
potential when computed in perturbation theory, on the other side its
direct lattice QCD computation suffers from poor convergence.  The convergence
can be improved by using gradient flow, where the gauge fields in the operator
definition of a given quantity are replaced by flowed fields at flow time $t$,
which effectively smear the gauge fields over a distance of order $\sqrt{t}$,
while they reduce to the QCD fields in the limit $t \to 0$.  Based on our
next-to-leading order calculation, we explore the properties of the static
force for arbitrary values of $t$, as well as in the $t \to 0$ limit, which may
be useful for lattice QCD studies.  }

\begin{document} \maketitle \flushbottom

\section{Introduction} \label{sec:intro}

The potential between a static quark and a static antiquark encodes important
information about the QCD interaction for a wide range of
distances~\cite{Wilson:1974sk, Susskind:1976pi, Fischler:1977yf, Brown:1979ya}.
While at long distances, the static potential exhibits a confining behavior,
the short-distance behavior can be directly compared with perturbative QCD
calculations~\cite{Fischler:1977yf,Schroder:1998vy,Brambilla:1999qa,Anzai:2009tm,Smirnov:2009fh}.
This has been proved useful in the extraction of the strong coupling
constant, $\alpha_s$, from lattice QCD computations of the static potential.
See
refs.~\cite{Karbstein:2014bsa,Bazavov:2014soa,Karbstein:2018mzo,Takaura:2018vcy,Bazavov:2019qoo,Ayala:2020odx}
for some recent determinations.

The perturbative QCD calculation of the static potential in dimensional
regularization  is affected by a renormalon of order
$\Lambda_{\rm QCD}$, which may
be absorbed in an overall constant shift~\cite{Pineda:1998id, Hoang:1998nz}.
Analogously, in lattice regularization, there is a linear divergence that is
proportional to the inverse of the lattice spacing.  
Because a constant shift in the potential has no
physical significance, it is preferable to directly probe the slope of the
static potential. 

The static force, which is defined by the spatial derivative of the static
potential, carries the essential information that determines the slope of the
static potential.  It  does not depend on the
constant shift in the potential 
which makes it convenient for comparing lattice with
perturbative studies~\cite{Necco:2001xg,Necco:2001gh,Pineda:2002se}.

The force can be computed from the finite differences of the lattice data of
the static potential.  This works well if the available data are dense, like in
the case of quenched lattice data~\cite{Necco:2001xg}.  In the case of full QCD
lattice studies, however, data at short distances are still sparse, and the
computation of the force from their finite differences leads to large
uncertainties~\cite{Bazavov:2014soa}.  To overcome this problem,
in~\cite{Vairo:2015vgb, Vairo:2016pxb} it has been suggested to compute the
force directly from a Wilson loop with a chromoelectric field insertion in it.
An exploratory study in lattice QCD of such a Wilson loop has been carried out
in ref.~\cite{Brambilla:2021wqs}. 

In ref.~\cite{Brambilla:2021wqs}, it has been found, however, that direct
lattice QCD calculations of the static force exhibit sizable discretization
errors and the convergence to the continuum limit is rather slow.  This poor
convergence may be understood from the convergence of the Fourier transform of
the perturbative QCD calculation in momentum space.  At tree level, the static
potential is given by the Fourier transform with respect to the spatial
momentum $\bm{q}$ of a function proportional to $1/\bm{q}^2$, which, after
integrating over the angles of $\bm{q}$, leads to an integral over $|\bm{q}|$
whose integrand decreases like $1/|\bm{q}|$ at large $|\bm{q}|$.  On the other
hand, the static force leads to an integral over $|\bm{q}|$ whose integrand
does not decrease with increasing $|\bm{q}|$.  As the lattice regularization
has the effect of introducing a momentum-space cutoff of the order of the
inverse of the lattice spacing, we can expect poor convergence to
the continuum limit when computing the static force directly in lattice QCD. 

The gradient-flow formulation has proven useful in lattice QCD calculations of
correlation functions and local operator matrix
elements~\cite{Narayanan:2006rf, Luscher:2009eq, Luscher:2010iy,Luscher:2011bx,
Luscher:2013vga, Borsanyi:2012zs, Suzuki:2013gza,Makino:2014taa,
Harlander:2018zpi}.  In gradient flow, the gauge fields in the operator
definitions of matrix elements are replaced by flowed fields that
depend on the spacetime coordinate and the flow time $t$.  The
flowed fields reduce to the bare gauge fields at $t=0$.  At tree level in
perturbation theory, the flowed fields come with a factor $e^{- q^2 t}$ for
every momentum-space gauge field with momentum $q$ in the operator definition.
If we compute the static force in gradient flow, the $e^{- q^2 t}$ factors will
make the integrand of the Fourier transform decrease faster than any power of
$\bm{q}$ at large $|\bm{q}|$.  If this behavior is kept unspoiled beyond tree
level, we expect that the poor convergence of the lattice QCD calculation of
the static force will be greatly improved by using gradient flow.

In this work, we compute the static force in gradient flow in perturbation
theory at next-to-leading order in the strong coupling.  This calculation is
significant in two aspects.  First, we examine the convergence of the Fourier
transform explicitly beyond tree level.  Second, we examine the dependence on
the flow time $t$, and in particular the behavior in the limit $t \to 0$, that
may be useful when extrapolating to QCD from lattice
calculations done in gradient flow. 

The paper is organized as follows.  In section~\ref{sec:defs}, we
define the static force and introduce the flowed fields in gradient flow.  We
compute the static force explicitly through next-to-leading order accuracy in
the strong coupling in section~\ref{sec:diags}.  Integral representations of
some coefficients are given in appendix~\ref{sec:oneloopfintable}.  We conclude
in section~\ref{sec:summary}.

\section{\boldmath Definitions and conventions} \label{sec:defs}

\begin{figure}[ht] \begin{center}
\includegraphics[width=0.4\textwidth]{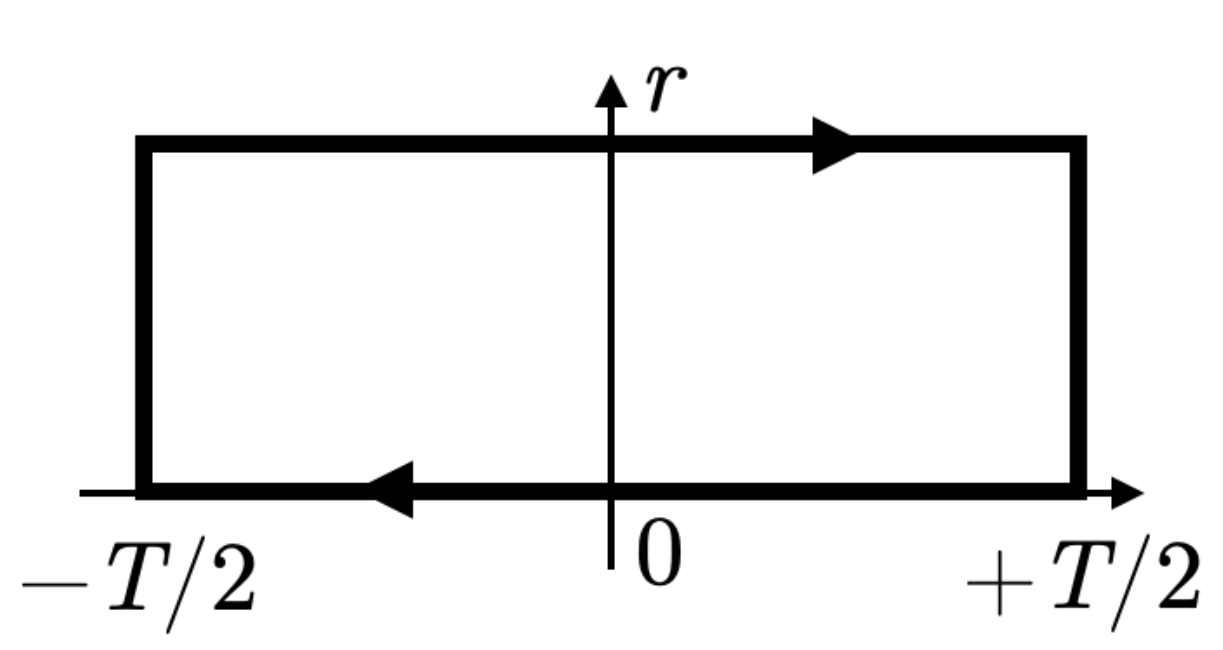} \caption{\label{fig:path} The
closed contour that defines the Wilson loop $W_{r \times T}$ in
eq.~\eqref{eq:wloop}.  This is a rectangle with spatial extension $r$ and
temporal extension $T$.  The horizontal axis represents time, and the vertical
axis is a spatial coordinate in an arbitrary direction.  } \end{center}
\end{figure}

We define the static potential in Euclidean QCD as~\cite{Wilson:1974sk,
Susskind:1976pi, Fischler:1977yf, Brown:1979ya} \begin{equation} V(r) = -
\lim_{T \to \infty} \frac{1}{T} \log \langle W_{r \times T} \rangle,
\end{equation} where $W_{r \times T}$ is a Wilson loop with temporal and
spatial extension $T$ and $r$, respectively, and $\langle \cdots \rangle$ is
the color-normalized time-ordered vacuum expectation value \begin{equation}
\langle \cdots \rangle = \frac{\langle 0 | {\cal T} \cdots | 0 \rangle}
{\langle 0 | {\rm tr}_{\rm color} {\bf 1}_{\rm c} | 0 \rangle}, \end{equation} with
${\cal T}$ the time ordering, $|0\rangle$ the QCD vacuum, and ${\rm tr}_{\rm
color} {\bf 1}_{\rm c}  = N_c$  the number of colors.  An explicit
expression for the Wilson loop $W_{r \times T}$ is \begin{equation}
\label{eq:wloop} W_{r \times T} = {\rm tr}_{\rm color} P \exp \left[ i g
\oint_C dz^\mu A_\mu (z) \right], \end{equation} where $P$ stands for the path
ordering of the color matrices, $A_\mu$ is the bare gluon field, $g$ is the
bare strong coupling, and $C$ is the closed contour shown in
figure~\ref{fig:path}. 

Following refs.~\cite{Brambilla:2000gk,Vairo:2015vgb,
Vairo:2016pxb,Brambilla:2019zqc,Brambilla:2021wqs}, we define the QCD static
force by the spatial derivative of $V(r)$ as \begin{equation}
\label{eq:forcedef} F(r) \equiv \frac{\partial}{\partial r} V(r) = -i \lim_{T
\to \infty} \frac{1}{T} \frac{\int_{-T/2}^{+T/2} dx_0 \,\langle W_{r \times T}
\hat{\bm{r}} \cdot g \bm{E}(x_0, \bm{r}) \rangle}{\langle W_{r \times T}
\rangle}, \end{equation} where in the second equality, the chromoelectric field
$g E_i = g F_{i0}$ is inserted into the Wilson loop at spacetime point
$(x_0,\bm{r})$; $F_{\mu \nu} = \partial_\mu A_\nu - \partial_\nu A_\mu - i
g[A_\mu, A_\nu]$ is the gluon field-strength tensor.  The perturbative QCD
expressions for $F(r)$ can be obtained from $V(r)$ by differentiating with
respect to $r$. 

The static force in gradient flow, $F(r;t)$, can be defined by replacing the
gluon fields $g A_\mu(x)$ by the flowed fields $B_\mu(x;t)$, where $B_\mu$ is
defined through the flow equation~\cite{Luscher:2010iy, Luscher:2011bx}
\begin{eqnarray} \frac{\partial}{\partial t} B_\mu (x;t) &=& D_\nu G_{\nu \mu}
+ \lambda D_\mu \partial_\nu B_\nu, \\ G_{\mu \nu} &=& \partial_\mu B_\nu -
\partial_\nu B_\mu + [B_\mu, B_\nu], \quad D_\mu = \partial_\mu + [B_\mu,
\cdot], \end{eqnarray} with the initial condition \begin{equation} B_\mu
(x;t=0) = g A_\mu(x).  \end{equation} Here, $\lambda$ is an arbitrary constant,
and the flow time $t$ is a variable of mass dimension $-2$.  Due to the initial
condition, the gradient-flow static force $F(r;t)$ reduces to $F(r)$ in the
limit $t \to 0$.  In perturbation theory, the flowed field in momentum space
$\tilde{B}_\mu$ can be written in terms of the usual momentum-space gluon field
$g \tilde{A}_\mu$ by solving iteratively the flow equation.  Since the
iterative solution takes a particularly simple form at $\lambda =1$, we set
$\lambda=1$ in our calculation.  In this case, we have~\cite{Luscher:2011bx}
\begin{eqnarray} \tilde{B}_\mu^a (p;t) &=& e^{-p^2 t} g \tilde{A}_\mu^a (p)
\nonumber \\ && + \int_0^t ds \, e^{-(t-s)p^2} \sum_{n=2}^3 \frac{1}{n!} \int
\frac{d^dq_1}{(2 \pi)^d} \cdots \int \frac{d^dq_n}{(2 \pi)^d} \delta (p+ q_1 +
\cdots + q_n) \nonumber \\ && \times X^{(n,0)} (p,q_1, \ldots, q_n)^{a b_1
\ldots b_n}_{\mu \nu_1 \ldots \nu_n} \tilde{B}_{\nu_1}^{b_1} (-q_1;s) \cdots
\tilde{B}_{\nu_n}^{b_n} (-q_n;s), \end{eqnarray} where $d$ is the number of
spacetime dimensions, and \begin{eqnarray} X^{(2,0)} (p,q_1, q_2)^{a b_1
b_2}_{\mu \nu_1 \nu_2} &=& i f^{a b_1 b_2} [ (q_2-q_1)_\mu \delta_{\nu_1 \nu_2}
+ 2 q_1{}_{\nu_2} \delta_{\mu \nu_1} - 2 q_2{}_{\nu_1} \delta_{\mu \nu_2} ], \\
X^{(3,0)} (p,q_1, q_2, q_3)^{a b_1 b_2 b_3}_{\mu \nu_1 \nu_2 \nu_3} &=& f^{a
b_1 c} f^{b_2 b_3 c} (\delta_{\mu \nu_3} \delta_{\nu_1 \nu_2} - \delta_{\mu
\nu_2} \delta_{\nu_1 \nu_3} ) \nonumber \\ && + f^{a b_3 c} f^{b_1 b_2 c}
(\delta_{\mu \nu_2} \delta_{\nu_1 \nu_3} - \delta_{\mu \nu_1} \delta_{\nu_2
\nu_3} ) \nonumber \\ && + f^{a b_2 c} f^{b_3 b_1 c} (\delta_{\mu \nu_1}
\delta_{\nu_2 \nu_3} - \delta_{\mu \nu_3} \delta_{\nu_1 \nu_2} ).
\end{eqnarray} We refer to the factor $e^{-(t-s)p^2}$ as the propagator of the
flow line from flow time $s$ to $t$, and to $X^{(n,0)}$ as flow vertices. 

For perturbative calculations, it is advantageous to first compute the static
potential in gradient flow, $V(r;t)$, by replacing the gluon fields $g
A_\mu(x)$ by the flowed fields $B_\mu(x;t)$ in the definition of the Wilson
loop, and then differentiate with respect to $r$ to find the static force
$F(r;t)$.  This is equivalent to computing $F(r;t)$ directly from the
expression involving the chromoelectric field, as the second equality in
eq.~\eqref{eq:forcedef} remains valid for nonzero $t$.  The perturbative QCD
calculation of the static potential can be further simplified by choosing a
gauge where the contributions from the spatial-direction Wilson lines at the
times $\pm T/2$ vanish in the limit $T \to \infty$.  It has been shown that in
Feynman gauge, the contributions from the gluon fields at the times $\pm T/2$
can indeed be neglected in computing the static
potential~\cite{Schroder:1999sg}.  Hence, we will employ the Feynman gauge and
consider only contributions from the temporal Wilson lines in the calculation
of the static potential. 

A simple set of Feynman rules can be found by going to momentum space and
setting $T \to \infty$~\cite{Schroder:1999sg}.  The momentum-space potential
$\tilde{V}(\bm{q})$ is related to the position-space counterpart by
\begin{equation} V(r) = \int \frac{d^3\bm{q}}{(2 \pi)^3} \tilde{V}(\bm{q}) e^{i
\bm{q} \cdot \bm{r}}.  \end{equation} The same relation holds for the static
potential in gradient flow.  The momentum-space Feynman rules for the
positive-time-direction temporal Wilson line are equivalent to the Feynman
rules for a static quark field in heavy-quark effective theory (HQET); whereas,
the negative-time-direction temporal Wilson line corresponds to a static
antiquark field in HQET~\cite{Schroder:1999sg, Eichten:1989zv}.  Since our goal
is to compute the static force, we neglect contributions to $\tilde{V}(\bm{q})$
with support only at vanishing $\bm{q}$, such as the heavy quark/antiquark
self-energy diagrams:  such contributions correspond to
$r$-independent constants that do not contribute to the static force.

Loop corrections to the momentum-space potential $V(\bm{q})$
involve both ultraviolet (UV) and infrared (IR) divergences, which require
regularization.  We regularize the divergences using dimensional regularization
in $d=4-2\epsilon$ spacetime dimensions.  While the IR divergences cancel in
the sum of all Feynman diagrams up to two
loops~\cite{Brambilla:1999qa,Brambilla:1999xf}, the UV divergences must be
removed by the renormalization of $\alpha_s = g^2/(4 \pi)$.  We renormalize the
strong coupling in the $\overline{\rm MS}$ scheme.  We associate a factor of $[
\mu^2 e^{\gamma_{\rm E}}/(4 \pi)]^\epsilon$ to every loop
integral, where $\mu$ is the $\overline{\rm MS}$ scale and $\gamma_{\rm E}$ is
the Euler--Mascheroni constant, so that renormalization in the $\overline{\rm
MS}$ scheme is carried out by simply subtracting the poles in $\epsilon$.

\section{\boldmath Computation of the static force} \label{sec:diags}

In this section we discuss the momentum-space calculation of the static
potential in gradient flow through next-to-leading order in $\alpha_s$, from
which we compute the static force in position space.

\subsection{\boldmath Leading order}

At order $g^0$, we have $\langle W_{r \times T} \rangle=1 + O(g^2)$, whose
contribution vanishes in $V(r;t)$.  The leading nontrivial contribution to
$V(r;t)$ occurs at order $g^2$.  At this order, the momentum-space potential is
given by \begin{equation} - \tilde{V}(\bm{q};t) = \frac{g^2 C_F}{\bm{q}^2}
e^{-2 \bm{q}^2 t} + O(\alpha_s^2), \end{equation} where $C_F = (N_c^2-1)/(2
N_c)$.  The factor $e^{-2\bm{q}^2 t}$ comes from the insertion of a flowed
$B_\mu(x;t)$ field on each of the two temporal Wilson lines.  As we have argued
in the previous section, we neglect the diagrams that contribute only at
$\bm{q}=\bm{0}$, such as the heavy quark/antiquark self-energy diagrams,
because they do not contribute to the static force.

\subsection{\boldmath Next-to-leading order} \label{sec:nlo}

\begin{figure}[ht] \begin{center}
\includegraphics[width=0.95\textwidth]{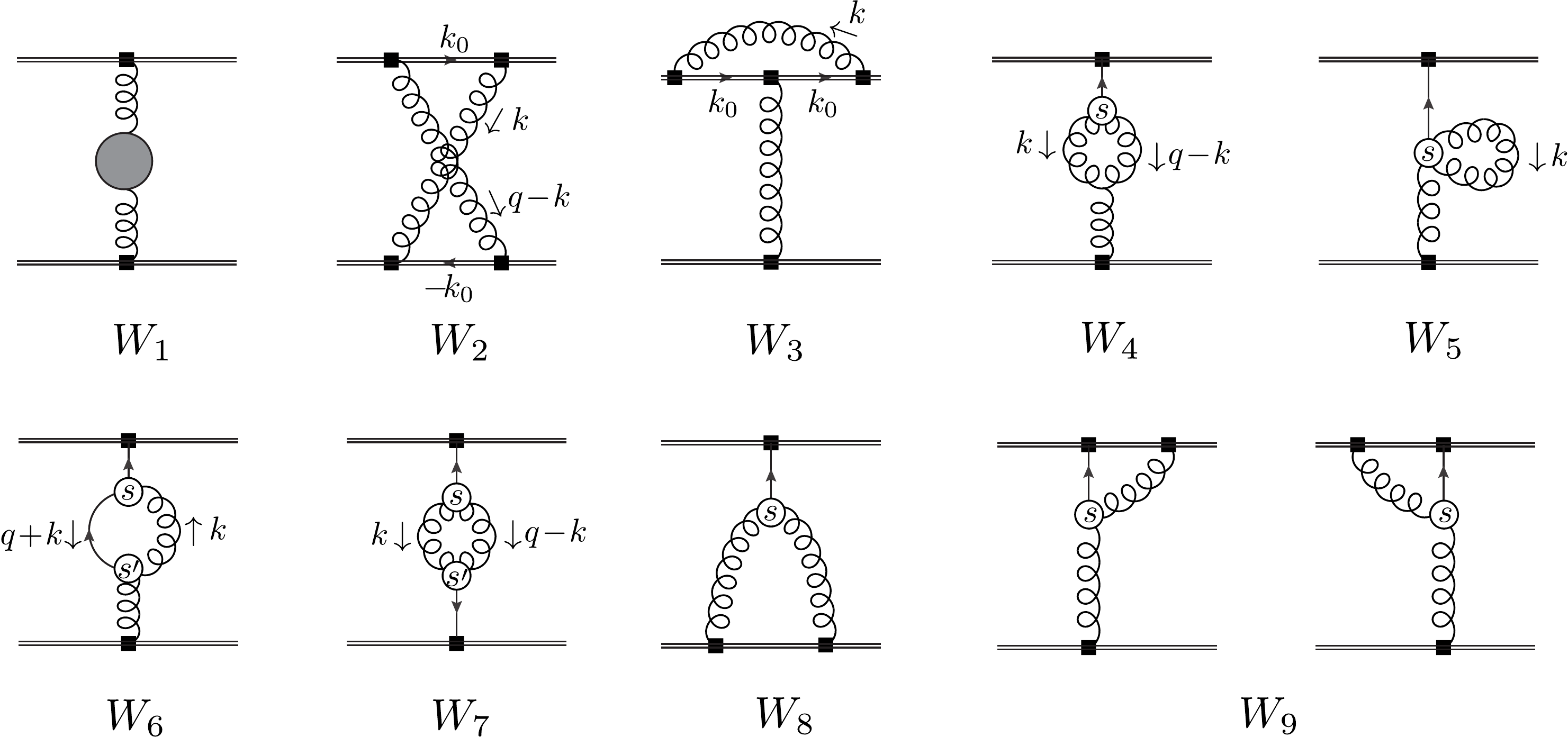}
\caption{\label{fig:potential_flow} Feynman diagrams for the static potential
at next-to-leading order in $\alpha_s$.  The double lines represent temporal
Wilson lines, curly lines are gluons, and solid lines are flow lines.  The
arrows on flow lines are in the direction of the greater flow time.  Filled
squares represent the flowed $B_\mu$ fields at flow time $t$, and open circles
are flow vertices $X^{(n,0)}$.  The blob in $W_1$ represents the gluon vacuum
polarization.  } \end{center} \end{figure}

The Feynman diagrams for the static potential at next-to-leading order (NLO) in
$\alpha_s$ (order $\alpha_s^2$) are shown in figure~\ref{fig:potential_flow}.
We include all combinatorial factors in $W_n$, so that the sum of all diagrams
is simply given by $\displaystyle \sum_{n=1}^9 W_n$.  Since we work in Feynman
gauge, we do not consider diagrams that vanish in this gauge, such as the
triple gluon vertex diagram.  Diagrams $W_1$--$W_3$ are the same as in QCD,
except that in gradient flow, the gluon fields are replaced by the
flowed fields $B_\mu$ at flow time $t$.  The remaining diagrams come from
contributions to the flowed fields $B_\mu$ beyond leading order in the
iterative solution of the flow equation, so that they involve flow lines, which
are represented by solid lines.  The diagrams $W_4$--$W_7$ correspond to
one-loop self-energy corrections to the flowed fields $B_\mu$; it has been
shown in  ref.~\cite{Luscher:2011bx} that the self-energy diagrams
corresponding to $W_4$--$W_6$ are UV divergent, while $W_7$ is finite. 

In the computation of the NLO diagrams, we neglect the contributions that
correspond to products of order-$\alpha_s$ contributions to $V(r)$, which arise
from the perturbative expansion of $\langle W_{r \times T} \rangle  \overset{T
\to \infty}{=} e^{-V(r) T}$ beyond linear order.  At order $\alpha_s^2$, the
contribution to $V(r)$ can be extracted by writing the color factors as linear
combinations of $C_F^2$ and $C_F C_A$, and discarding the terms proportional to
$C_F^2$, because the $C_F^2$ terms correspond to the $[-V(r) T]^2/2$ term in
the perturbative expansion of $e^{-V(r) T}$. Here, $C_A = N_c$. 

We now discuss one by one the computation of the diagrams in
figure~\ref{fig:potential_flow}.  Diagram $W_1$ comes from the gluon vacuum
polarization at one loop.  In Feynman gauge, $W_1$ is given by \begin{equation}
W_1 = \frac{g^4 C_F }{16 \pi^2 \bm{q}^2} e^{-2 \bm{q}^2 t} \left[ \frac{31}{9}
C_A - \frac{10}{9} n_f + \left( \frac{5}{3} C_A - \frac{2}{3} n_f \right)
\left( \frac{1}{\epsilon_{\rm UV}} + \log (\mu^2/\bm{q}^2) \right) \right],
\end{equation} where $n_f$ is the number of massless quark flavors.  
We use the label UV to indicate the ultraviolet origin of the
$1/\epsilon$ pole.

Diagram $W_2$ is given by \begin{eqnarray} W_2 &=& \frac{1}{2} g^4 C_F C_A \int
\frac{d^dk}{(2 \pi)^d} \frac{ e^{- 2 t [k^2+(q-k)^2]} }{k^2 (q-k)^2 ( k_0 + i
\varepsilon)^2} \nonumber \\ &=& -\frac{1}{2} g^4 C_F C_A \frac{e^{-2 \bm{q}^2
t}}{\bm{q}^{2+2 \epsilon}} \frac{({\mu^2 e^{\gamma_{\rm E}}})^\epsilon}{8
\pi^2} \int_0^\infty dx \int_0^\infty dy \, \frac{ \exp \left( \frac{ 4
\bar{t}^2-x y}{4 \bar{t}+x+y} \right) }{(x+y+4\bar{t})^{1-\epsilon}},
\end{eqnarray} where we have used Schwinger parametrization, we have integrated
over $k$ to obtain the parameter integral in the last line, and set
$\bar{t} \equiv \bm{q}^2 t$.  The parameter integral is IR divergent for
$\epsilon =0$, which can be seen from the behavior of the integrand at $x \to
\infty$ or $y \to \infty$.  We employ sector decomposition~\cite{Hepp:1966eg,
Binoth:2000ps, Binoth:2003ak,Heinrich:2008si} to compute the parameter
integral, which provides a way to separate the divergent and finite
contributions.  The finite contribution is obtained in the form of a parameter
integral that is finite at $\epsilon = 0$, so that we can expand the integrand
in powers of $\epsilon$.  We change the integration variables 
according to $x=\bar{t} x_1/(1-x_1)$ and $y=\bar{t} x_2/(1-x_2)$, so that
the region of integration is a unit square given by $0 < x_1 < 1$ and $0 < x_2
<1$.  The region of integration is then divided into the sectors $x_1 < x_2$
and $x_1 > x_2$; the contribution from each sector can again be expressed as an
integral over a unit square by rescaling the integration variables. We obtain
\begin{equation} W_2 = g^4 C_F C_A \frac{e^{-2 \bm{q}^2 t}}{16 \pi^2 \bm{q}^2}
\left[ 2 \left( \frac{1}{\epsilon_{\rm IR}} + \log(\mu^2 /\bm{q}^2) \right) +
W_2^{F} (\bar{t}) + O(\epsilon) \right], \end{equation} where we use the label
IR to indicate the infrared origin of the $1/\epsilon$ pole, and  $W_2^{F}
(\bar{t})$ is the finite piece given by \begin{equation} W_2^{F} (\bar{t}) = 2
\left[ e^{2 \bar{t}} \left( \text{Ei}(-\bar{t})-2 \, \text{Ei}(-2 \bar{t})
\right) -\text{Ei}(\bar{t})+2 \log 2 + 2 \log \bar{t}+2 \gamma_{\rm E} \right]
.  \end{equation}  The function $\text{Ei} (z)$ is the
exponential integral defined by \begin{equation} \text{Ei}(x) =
\int_{-x}^\infty \frac{e^{-t}}{t} dt.  \end{equation}

Diagram $W_3$ can be computed in a similar way. We have \begin{equation} W_3 =
- g^4 C_F C_A \frac{e^{-2 \bm{q}^2 t}}{\bm{q}^2} \int \frac{d^dk}{(2 \pi)^d}
\frac{1}{k^2 (k_0 +i \varepsilon)^2} e^{-2t k^2}.  \end{equation} We use
Schwinger parametrization to combine the $k^2$ in the denominator and the
exponent.  We obtain \begin{eqnarray} W_3 &=& - g^4 C_F C_A \frac{e^{-2
\bm{q}^2 t}}{\bm{q}^2} \int_0^\infty dx \int \frac{d^dk}{(2 \pi)^d}
\frac{1}{(k_0 +i \varepsilon)^2} e^{-(x+2t) k^2} \nonumber \\ &=& \frac{(\mu^2
e^{\gamma_{\rm E}})^\epsilon}{8 \pi^2} g^4 C_F C_A \frac{e^{-2 \bm{q}^2
t}}{\bm{q}^2} \int_0^\infty \frac{dx}{(x+2 t)^{1-\epsilon}} \nonumber \\ &=&
g^4 C_F C_A \frac{e^{-2 \bm{q}^2 t}}{16 \pi^2 \bm{q}^2} \left[ - 2 \left(
\frac{1}{\epsilon_{\rm IR}} + \log (\mu^2/\bm{q}^2) \right) + W_3^F(\bar{t})  +
O(\epsilon) \right], \end{eqnarray} where \begin{equation} W_3^F(\bar{t}) = - 2
[\log ( 2 \bar{t}) +\gamma_{\rm E} ].  \end{equation} The pole in $\epsilon$
comes from the divergence of the parameter integral at large $x$, which
corresponds to the vanishing of $k^2$.  Hence, the pole in $W_3$ is an IR pole.
We note that the IR poles cancel in $W_2 + W_3$, similarly to the QCD
case~\cite{Fischler:1977yf}. 

The computation of the UV-divergent diagrams $W_4$--$W_6$ can be done in a
similar way. We have \begin{eqnarray} W_4 &=& \frac{2 g^4 C_A C_F}{\bm{q}^2}
\int_0^t ds \, e^{- (2 t-s) q^2} \int \frac{d^dk}{(2 \pi)^d} e^{-s
(k^2+(q-k)^2)} \frac{ 2 (d-2) k_0^2 + q^2+ 2 k^2 }{k^2 (q-k)^2}, \\ W_5 &=&
-\frac{2 (d-1) g^4 C_A C_F}{\bm{q}^2} e^{-2 \bm{q}^2 t} \int_0^t ds\, \int
\frac{d^dk}{(2 \pi)^d} \frac{1}{k^2} e^{-2 s k^2}, \\ W_6 &=& \frac{4 g^4 C_A
C_F }{\bm{q}^2} e^{-2 \bm{q}^2 t} \int_0^t ds \int_0^s ds' \int \frac{d^dk}{(2
\pi)^d} e^{ s q^2 -(s-s') (q+k)^2 -(s+s') k^2 - s' q^2} \nonumber \\ &&
\hspace{35ex} \times \frac{2 (d-2) k_0^2 + k^2 + k \cdot q + 2 q^2 }{k^2}.
\end{eqnarray} We use again Schwinger parametrization to exponentiate the gluon
propagator denominators.  After integrating over $k$, we obtain expressions
that are integrals over the Schwinger parameters and the flow times $s$ and
$s'$.  These integrals can then be reexpressed as integrals over a unit
hypercube, in a way similar to what we have done for the computation of $W_2$,
and by rescaling the flow times $s$ and $s'$.  We then use sector decomposition
to extract the UV poles of the parameter integrals and obtain the following
expressions: \begin{eqnarray} W_4 &=& g^4 C_A C_F \frac{e^{-2 \bm{q}^2 t} }{16
\pi^2 \bm{q}^2} \left[ 3 \left( \frac{1}{\epsilon_{\rm UV}} + \log (\mu^2
/\bm{q}^2) \right) + W_4^F(\bar{t}) + O(\epsilon) \right], \\ W_5 &=& g^4 C_A
C_F \frac{e^{-2 \bm{q}^2 t} }{16 \pi^2 \bm{q}^2} \left[ - 3
\left(\frac{1}{\epsilon_{\rm UV}} + \log (\mu^2 /\bm{q}^2) \right) +
W_5^F(\bar{t}) + O(\epsilon) \right], \\ W_6 &=& g^4 C_A C_F \frac{e^{-2
\bm{q}^2 t} }{16 \pi^2 \bm{q}^2} \left[ 2 \left( \frac{1}{\epsilon_{\rm UV}} +
\log (\mu^2 /\bm{q}^2) \right) + W_6^F(\bar{t}) + O(\epsilon) \right],
\end{eqnarray} where the terms $W_4^F(\bar{t})$, $W_5^F (\bar{t})$, and
$W_6^F(\bar{t})$ can be written as integrals over a unit hypercube that are
finite at $d=4$.  The analytical expression for $W_5^F(\bar{t})$ is given by
\begin{equation} W_5^F (\bar{t}) = - 1- 3 \gamma_{\rm E} - 3 \log (2 \bar{t}) .
\end{equation} We have not found analytical expressions for $W_4^F(\bar{t})$
and $W_6^F(\bar{t})$.  We show $W_4^F(\bar{t})$ and $W_6^F(\bar{t})$ as
integrals over a unit hypercube in appendix~\ref{sec:oneloopfintable}.  The UV
poles of $W_4$ and $W_5+W_6$ agree with the results in
ref.~\cite{Luscher:2011bx}.

The remaining diagrams $W_7$--$W_9$ yield \begin{eqnarray} W_7 &=& 2 g^4 C_A
C_F \int_0^t ds \int_0^t ds' \int \frac{d^dk}{(2 \pi)^d} e^{-(s+s')
(k^2+(k-q)^2)-(t-s-s') q^2} \frac{(d-2) k_0^2 + 2 k^2}{k^2 (q-k)^2}, \quad\quad
\\ W_8 &=& 2 g^4 C_A C_F \int_0^t ds \, e^{- (t-s) q^2} \int \frac{d^dk}{(2
\pi)^d} e^{-s (k^2+(q-k)^2)} \frac{e^{-k^2 t}}{k^2} \frac{e^{-(q-k)^2
t}}{(q-k)^2}, \\ W_9 &=& - 2 g^4 C_A C_F \int \frac{d^dk}{(2 \pi)^d} \int_0^t
ds \, e^{- (t-s) (q-k)^2} e^{-s (k^2+q^2)} \frac{e^{-q^2 t}}{q^2} \frac{e^{-k^2
t}}{k^2}.  \end{eqnarray} Since these diagrams are finite, we can compute them
at $d=4$.  We again compute these diagrams by using Schwinger parametrization.
After integrating over $k$, we obtain \begin{eqnarray} W_7 &=& g^4 C_A C_F
\frac{e^{-2 \bm{q}^2 t}}{16 \pi^2 \bm{q}^2} W_7^F(\bar{t}) +O(\epsilon), \\ W_8
&=& g^4 C_A C_F \frac{e^{-2 \bm{q}^2 t}}{16 \pi^2 \bm{q}^2} W_8^F(\bar{t})
+O(\epsilon), \\ W_9 &=& g^4 C_A C_F \frac{e^{-2 \bm{q}^2 t}}{16 \pi^2
\bm{q}^2} W_9^F(\bar{t})  +O(\epsilon), \end{eqnarray} where $W_7^F(\bar{t})$,
$W_8^F(\bar{t})$, and $W_9^F(\bar{t})$ are parameter integrals over a unit
hypercube; explicit expressions can be found in
appendix~\ref{sec:oneloopfintable}.  The expressions for $W_8^F$ and $W_9^F$
can be further simplified into \begin{eqnarray} W_8^F (\bar{t}) &=& -2 \left\{
2 e^{2 \bar{t}} \text{Ei}\left(-2 \bar{t}\right)+e^{\bar{t}}
\left[\text{Ei}\left(-\frac{\bar{t}}{2}\right)-(e^{\bar{t}}+2)
\text{Ei}\left(-\bar{t}\right)\right]+\text{Ei}\left(\frac{\bar{t}}{2}\right)
-\text{Ei}\left(\bar{t}\right)\right\}, \quad \quad \\ W_9^F (\bar{t}) &=&
\frac{1}{\bar{t}} \left[ -\sqrt{2 \pi \bar{t}} \,
\text{erfi}\left(\sqrt{\bar{t}/2}\right) +2 e^{\bar{t}/2}-2 \right],
\end{eqnarray} where, $\text{erfi} (z) = -i \, \text{erf} (i z)$ is the
imaginary error function, and \begin{equation} \text{erf} (z) =
\frac{2}{\sqrt{\pi}} \int_0^z dt \, e^{-t^2} .  \end{equation}

As a cross check of our results, we have also evaluated the diagrams 
$W_2$--$W_9$ by using different parameterizations of the Feynman integrals and
evaluating them numerically, and found agreements in all cases. In the case of 
divergent diagrams, the use of different parameterizations leads to expressions
of divergent and finite integrands that differ nontrivially from what we
have obtained above, and so, the agreement in the sum of the divergent and
finite contributions to each diagram serves as an independent check of our 
results.

\subsection{\boldmath Results in momentum space} \label{res:mom}

The total one-loop correction to $-\tilde{V}(\bm{q};t)$ is given by
\begin{equation} \sum_{n=1}^9 W_n = \frac{4 \pi \alpha_s C_F e^{-2 \bm{q}^2 t}
}{\bm{q}^2} \frac{\alpha_s}{4 \pi} \bigg[ \beta_0 \left(\frac{1}{\epsilon_{\rm
UV} } + \log(\mu^2 /\bm{q}^2) \right) + a_1 + C_A \, W_{\rm NLO}^F (\bar{t}) +
O(\epsilon) \bigg], \end{equation} where $\displaystyle \beta_0 = \frac{11}{3}
C_A - \frac{2}{3} n_f$, and \begin{eqnarray} a_1 &=& \frac{31}{9} C_A -
\frac{10}{9} n_f, \\ W_{\rm NLO}^F (\bar{t}) &=& \sum_{n=2}^9 W_{n}^F (\bar{t})
.  \end{eqnarray} The UV pole is subtracted by renormalization of the strong
coupling.  In the $\overline{\rm MS}$ scheme, we find \begin{eqnarray}
\tilde{V}(\bm{q};t) &=& - \frac{4 \pi \alpha_s (\mu) C_F e^{-2 \bm{q}^2 t}
}{\bm{q}^2} \nonumber \\ && \times \bigg\{ 1+ \frac{\alpha_s (\mu)}{4 \pi}
\bigg[ \beta_0 \log(\mu^2 /\bm{q}^2) + a_1 + C_A \, W_{\rm NLO}^F (\bar{t})
\bigg]\bigg\} + O(\alpha_s^3), \end{eqnarray} where now $\alpha_s (\mu)$ is the
strong coupling constant in the $\overline{\rm MS}$ scheme, and $\mu$ is the
renormalization scale.  The term $a_1$ is the finite piece of the one-loop
correction to the static potential in regular QCD~\cite{Fischler:1977yf}, while
$C_A \, W_{\rm NLO}^F (\bar{t})$ is the extra finite term that appears in
gradient flow. 

\begin{figure}[ht] \begin{center}
\includegraphics[width=0.6\textwidth]{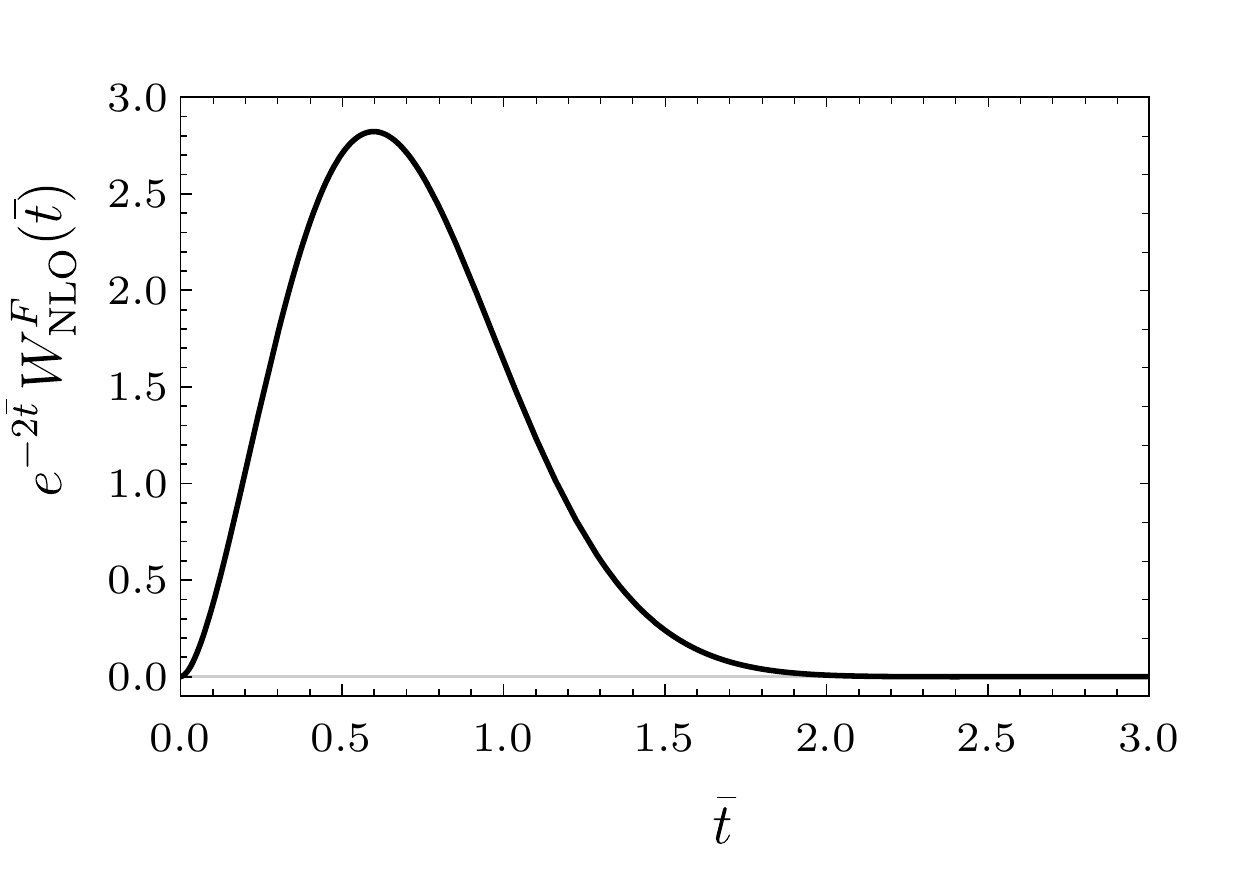}
\caption{\label{fig:WFNLOplot} The finite correction term $e^{-2 \bar{t}}
W_{\rm NLO}^F (\bar{t})$ as a function of $\bar{t} = \bm{q}^2t$.  }
\end{center} \end{figure}

Since we have not obtained analytical results for some of the one-loop
diagrams, we compute $W_{\rm NLO}^F (\bar{t})$ for arbitrary $\bar{t}$ by
numerically evaluating the parameter integrals given in the appendix.  As
$\bar{t}$ increases, $W_{\rm NLO}^F (\bar{t})$ diverges like
$e^{\bar{t}}/\bar{t}$.  However, when multiplied by the factor $e^{-2 \bm{q}^2
t} = e^{-2 \bar{t}}$, the extra finite term still decreases exponentially at
large $\bar{t}$.  We show the numerical results for $ e^{-2 \bar{t}} \, W_{\rm
NLO}^F (\bar{t})$ in figure~\ref{fig:WFNLOplot}. 

It is possible to show analytically that $W_{\rm NLO}^F (\bar{t})$ vanishes in
the limit $\bar{t} \to 0$, so that we recover the QCD result as expected.  From
the analytical expressions, we have the following expansions at $\bar{t}=0$:
\begin{eqnarray} W_2^F(\bar{t}) &=& -4 \bar{t} \left[ -1 + \gamma_{\rm E} +
\log (4 \bar{t}) \right] + O(\bar{t}^2), \\ W_8^F (\bar{t}) &=& -2 \bar{t} [-2
+ \gamma_{\rm E} + \log ( 8 \bar{t}) ] + O(\bar{t}^2), \\ W_9^F(\bar{t}) &=& -
1 - \frac{\bar{t}}{12}+ O(\bar{t}^2), \end{eqnarray} moreover we find the
following expansions by computing the parameter integrals near $t=0$:
\begin{eqnarray} W_4^F(\bar{t}) &=& 3 \log (2 \bar{t}) + 3 \gamma_{\rm E} +
\frac{5}{2} + \bar{t} \left[ - \frac{4}{3} \log(2 \bar{t}) - \frac{4}{3}
\gamma_{\rm E} + \frac{31}{9} \right] +O(\bar{t}^2), \\ W_6^F(\bar{t}) &=& 2
\log (2 \bar{t}) +2 \gamma_{\rm E} - \frac{1}{2} + \frac{145}{36} \bar{t}
+O(\bar{t}^2), \\ W_7^F(\bar{t}) &=& 5 \bar{t} \log 2 +O(\bar{t}^2).
\end{eqnarray} The $\bar{t} \log \bar{t}$ term in $W_4^F(\bar{t})$ arises from
divergences in the parameter integral when the integrand is expanded in powers
of $\bar{t}$; the coefficient of the $\bar{t} \log \bar{t}$ term can be
computed by isolating the divergent regions in the limit $\bar{t} \to 0$.  By
using the above expansions and the exact analytical expressions of
$W_3^F(\bar{t})$ and $W_5^F(\bar{t})$, we obtain \begin{equation}
\label{eq:WNLOF_exp} W_{\rm NLO}^F (\bar{t}) = \bar{t} \left( -\frac{22
\gamma_{\rm E} }{3}+\frac{277}{18}-\frac{31 \log 2}{3} - \frac{22}{3} \log
\bar{t} \right) + O(\bar{t}^2).  \end{equation} Since the $\bar{t} \log
\bar{t}$ terms do not cancel in $W_{\rm NLO}^F (\bar{t})$, we find that the
$t$-dependent one-loop finite term is not analytical at $t=0$, even though
$W_{\rm NLO}^F (\bar{t})$ vanishes in the limit $t \to 0$ as expected.  The
order-$\bar{t}$ term in eq.~\eqref{eq:WNLOF_exp} will be crucial in obtaining
the behavior of the position-space static force in the limit $t \to 0$ in the
next section.

\subsection{\boldmath Results in position space} \label{res:pos}

We now turn to position space to obtain \begin{equation} F(r;t) =
\frac{\partial}{\partial r} \int \frac{d^3\bm{q}}{(2 \pi)^3}
\tilde{V}(\bm{q};t) e^{i \bm{q} \cdot \bm{r}} = \frac{1}{r^2} \int_0^\infty d
|\bm{q}| \bm{q}^2 \frac{|\bm{q}| r \cos(|\bm{q}| r) - \sin(|\bm{q}| r)}{2 \pi^2
|\bm{q}|} \tilde{V}(\bm{q};t).  \end{equation} By using the fact that $r^2
F(r;t)$ is dimensionless, we define the following dimensionless quantities
\begin{eqnarray} {\cal F}_0 (r;t) &=& -\int_0^\infty d |\bm{q}| \bm{q}^2
\frac{|\bm{q}| r \cos(|\bm{q}| r) - \sin(|\bm{q}| r)}{2 \pi^2 |\bm{q}|} \frac{4
\pi e^{-2 \bm{q}^2 t}}{\bm{q}^2}, \\ {\cal F}_{\rm NLO}^L (r;t;\mu) &=&
-\int_0^\infty d |\bm{q}| \bm{q}^2 \frac{|\bm{q}| r \cos(|\bm{q}| r) -
\sin(|\bm{q}| r)}{2 \pi^2 |\bm{q}|} \frac{4 \pi e^{-2 \bm{q}^2 t}}{\bm{q}^2}
\log (\mu^2/\bm{q}^2), \\ {\cal F}_{\rm NLO}^F (r;t) &=& -\int_0^\infty d
|\bm{q}| \bm{q}^2 \frac{|\bm{q}| r \cos(|\bm{q}| r) - \sin(|\bm{q}| r)}{2 \pi^2
|\bm{q}|} \frac{4 \pi e^{-2 \bm{q}^2 t}}{\bm{q}^2} W_{\rm NLO}^F(\bar{t} =
\bm{q}^2 t), \end{eqnarray} so that \begin{eqnarray} F(r;t) &=&
\frac{\alpha_s(\mu) C_F}{r^2} \bigg[ \left( 1 + \frac{\alpha_s}{4 \pi} a_1
\right) {\cal F}_0 (r;t) \nonumber \\ && \hspace{12ex} + \frac{\alpha_s}{4 \pi}
\beta_0 {\cal F}_{\rm NLO}^L (r;t;\mu) + \frac{\alpha_s C_A}{4 \pi} {\cal
F}_{\rm NLO}^F (r;t) \bigg] + O(\alpha_s^3).  \end{eqnarray} The integrands for
${\cal F}_0 (r;t)$ and ${\cal F}_{\rm NLO}^L (r;t;\mu)$ involve the factor
$e^{-2 \bm{q}^2 t}$ that makes the integrands decrease exponentially at large
$|\bm{q}|$, so that the Fourier transforms converge rapidly.  In the case of
${\cal F}_{\rm NLO}^F (r;t)$, the convergence is slower, because $W_{\rm
NLO}^F(\bar{t})$ diverges exponentially at large $\bar{t}$.  However, since
$e^{-2 \bm{q}^2 t} W_{\rm NLO}^F(\bar{t} = \bm{q}^2 t)$ still decreases
exponentially at large $|\bm{q}|$, the integral in ${\cal F}_{\rm NLO}^F (r;t)$
still converges rapidly. 

\begin{figure}[ht] \begin{center}
\includegraphics[width=0.6\textwidth]{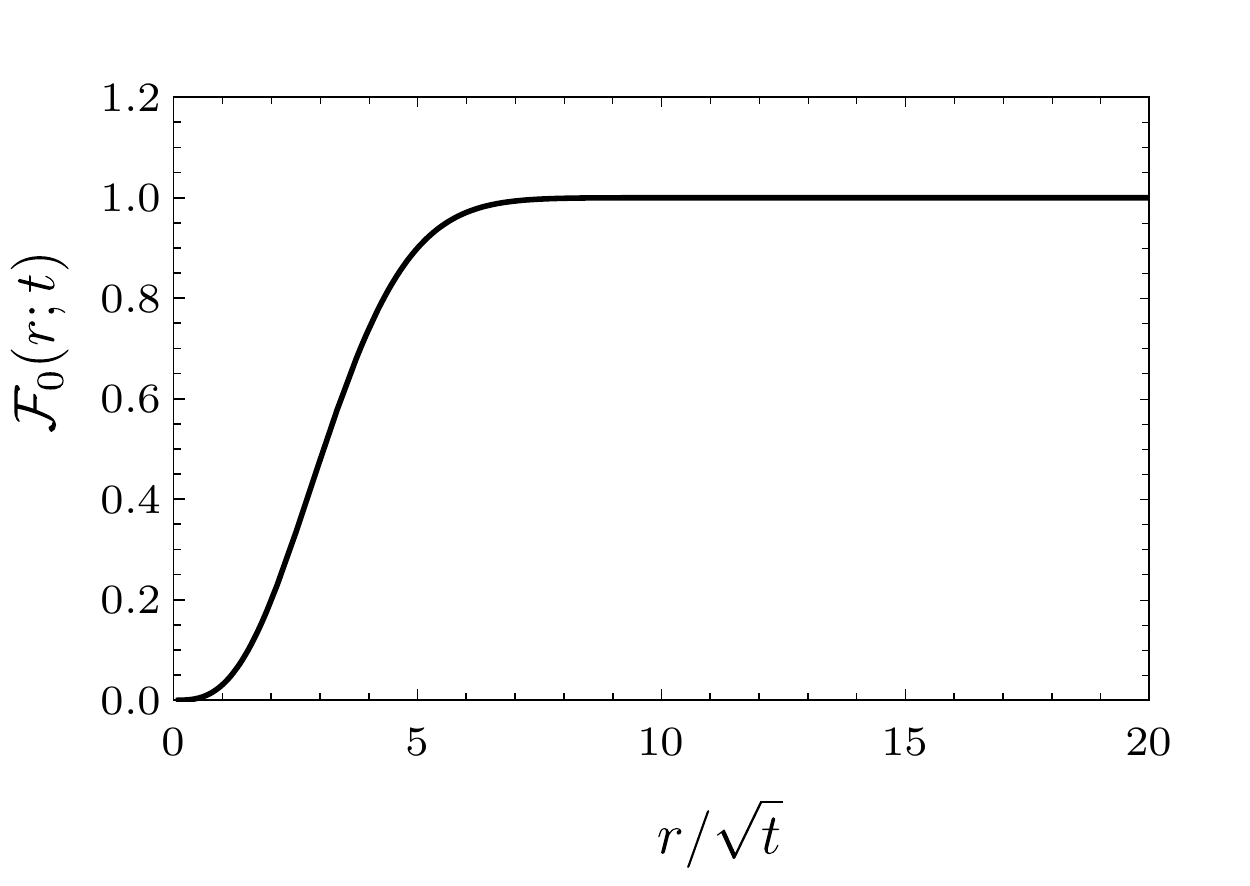} \caption{\label{fig:F0plot}
The factor ${\cal F}_0 (r;t)$ as a function
of $r/\sqrt{t}$.  } \end{center} \end{figure}

The quantities ${\cal F}_0 (r;t)$ and ${\cal F}_{\rm NLO}^L (r;t;\mu)$ can be
computed analytically: \begin{eqnarray} {\cal F}_0 (r;t) &=& \text{erf} \left(
\frac{r}{\sqrt{8 t}} \right) - \frac{r}{\sqrt{2 \pi t}} \exp \left( -
\frac{r^2}{8 t} \right) , \\ {\cal F}_{\rm NLO}^L (r;t;\mu) &=& \log (\mu^2
r^2) {\cal F}_0 (r;t) + \log \left( \frac{8 t}{r^2} e^{\gamma_{\rm E}} \right)
{\cal F}_0 (r;t)  \nonumber\\ && - \frac{r}{\sqrt{2 \pi t}} \left[
e^{-\frac{r^2}{8 t}} M^{(1,0,0)} \left(0,\frac{1}{2},\frac{r^2}{8 t} \right) +
M^{(1,0,0)} \left(\frac{1}{2},\frac{3}{2},-\frac{r^2}{8 t} \right) \right],
\end{eqnarray} where $M(a,b,z)$ is the confluent hypergeometric function
defined by \begin{equation} M (a,b,z) = \sum_{k=0}^\infty \frac{(a)_k}{(b)_k}
\frac{z^k}{k!}, \end{equation} with $(x)_k = \Gamma(x+k)/\Gamma(x)$, and
\begin{equation} M^{(1,0,0)} (a,b,z) = \frac{\partial}{\partial a} M (a,b,z).
\end{equation} We note that ${\cal F}_0 (r;t)$ is a function of $r/\sqrt{t}$
only.  This function vanishes at $r=0$, and $\displaystyle \lim_{t \to 0} {\cal
F}_0 (r;t) =1$.  We show ${\cal F}_0 (r;t)$ as a function of $r/\sqrt{t}$ in
figure~\ref{fig:F0plot}.

\begin{figure}[ht] \begin{center}
\includegraphics[width=0.6\textwidth]{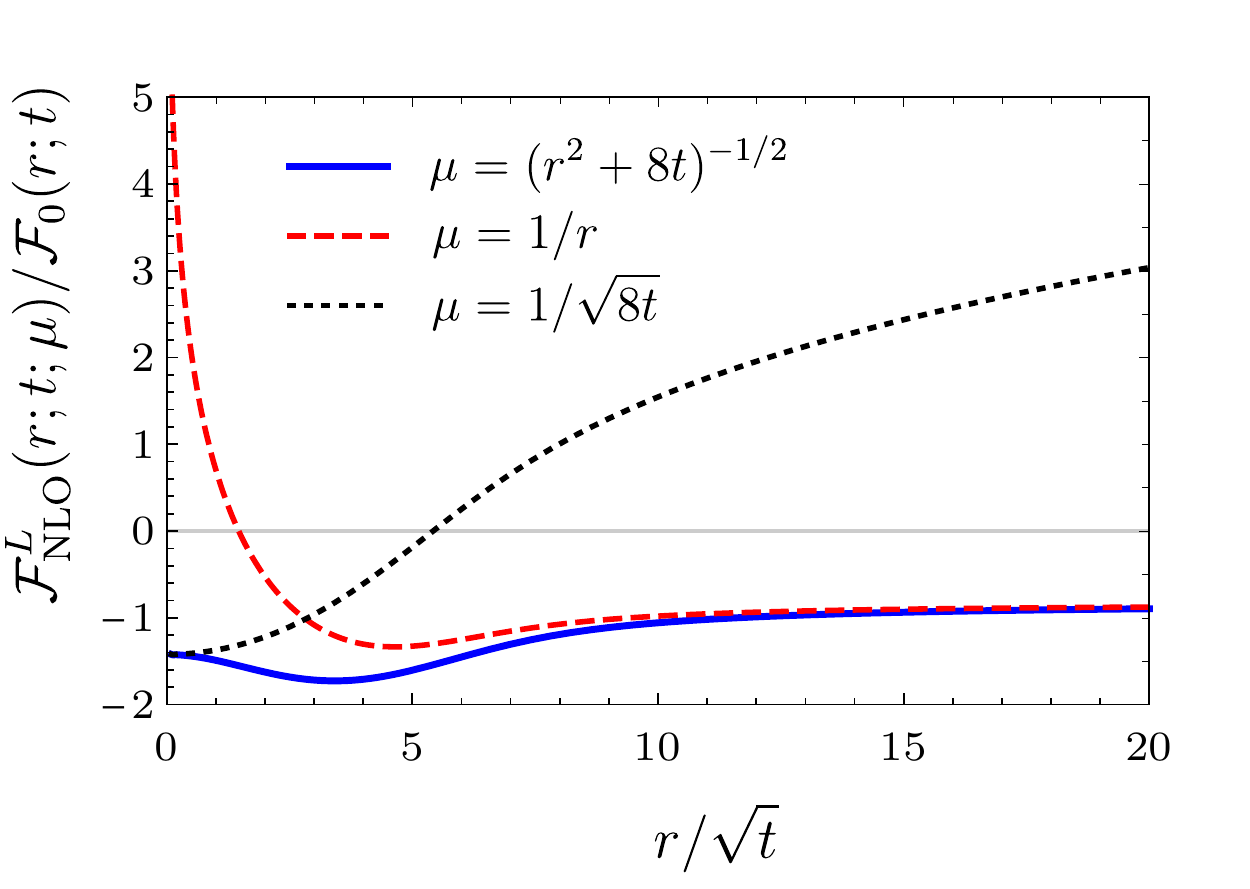} \caption{\label{fig:FLmudep}
The logarithmic correction factor ${\cal F}_{\rm NLO}^L (r;t;\mu)/{\cal
F}_0(r;t)$ for $\mu = ( r^2 + 8 t )^{-1/2}$, $\mu = 1/r$, and $\mu = 1/\sqrt{8
t}$ shown as a function of $r/\sqrt{t}$.  }
\end{center} \end{figure}

Combining the corrections $\log (\mu^2 r^2){\cal F}_0 (r;t)$ and $\log (8
t/r^2){\cal F}_0 (r;t)$ in ${\cal F}_{\rm NLO}^L (r;t;\mu)$ may suggest that
the logarithmic corrections in ${\cal F}_{\rm NLO}^L$ will be suppressed if we
set $\mu = 1/\sqrt{8 t}$.  This is indeed the case for $r^2/t \ll 1$, since in
this limit the hypergeometric functions in ${\cal F}_{\rm NLO}^L (r;t;\mu)$ are
analytic.  On the other hand, the contributions to ${\cal F}_{\rm NLO}^L
(r;t;\mu)$ other than $\log (\mu^2 r^2) {\cal F}_0 (r;t)$ assume the constant
value $2 (\gamma_{\rm E} -1)$ in the opposite limit $r^2 / t \to \infty$.
Hence, the relevant scale for the one-loop correction is $\mu = 1/r$ for $r^2/t
\gg 1$, while it is $\mu = 1/\sqrt{8 t}$ for $r^2/t \ll 1$.  We find that the
scale choice $\mu = ( r^2 + 8 t )^{-1/2}$ makes the logarithmic correction
factor ${\cal F}_{\rm NLO}^L (r;t;\mu)/{\cal F}_0(r;t)$ of order 1 for all
values of $r/\sqrt{t}$, in contrast to the choices $\mu=1/r$ or $\mu =
1/\sqrt{8 t}$.  This is shown in figure~\ref{fig:FLmudep}.  
Furthermore we have that ${\cal F}_{\rm NLO}^L(r;t;\mu)$ approaches the
limit $t\to 0$ linearly in $t$, in the form \begin{equation} {\cal F}_{\rm
NLO}^L(r;t;\mu) \approx \log(\mu^2 r^2) + 2 (\gamma_{\rm E} -1) - 12
\frac{t}{r^2}, \end{equation} which can be obtained from the asymptotic
expansion of $M(a,b,z)$ for large $z$ (see, for example,
ref.~\cite{NIST:DLMF}). 

\begin{figure}[ht] \begin{center}
\includegraphics[width=0.6\textwidth]{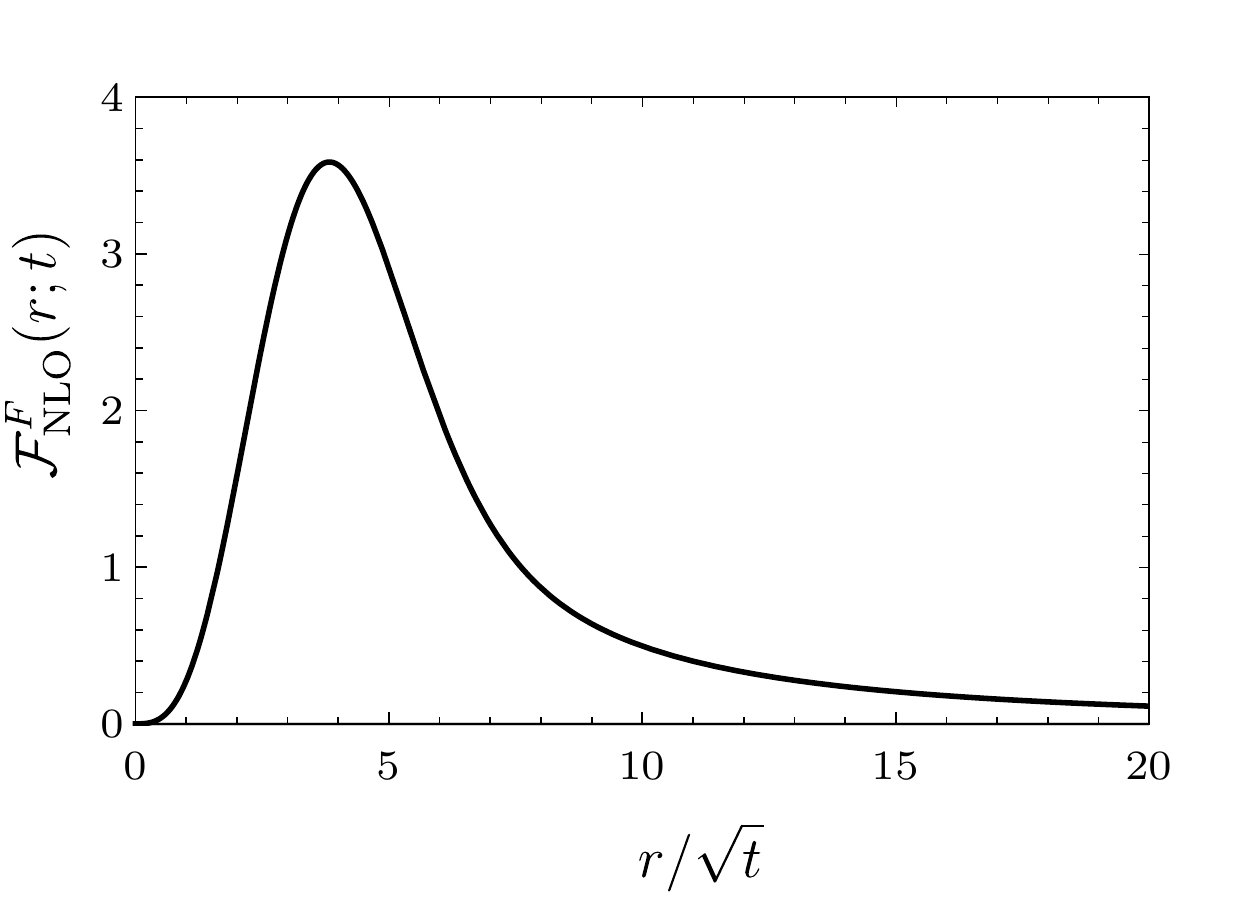}
\caption{\label{fig:FNLOplot} The finite one-loop correction ${\cal F}_{\rm
NLO}^F (r;t)$ as a function of $r/\sqrt{t}$.  } \end{center} \end{figure}

Now we discuss the finite correction term ${\cal F}_{\rm NLO}^F (r;t)$.  Since
${\cal F}_{\rm NLO}^F (r;t)$ is dimensionless, it is a function of $\xi =
r/\sqrt{t}$ only, and we can rewrite ${\cal F}_{\rm NLO}^F (r;t)$ as
\begin{equation} \label{eq:fnloalt} {\cal F}_{\rm NLO}^F (r;t) = -
\frac{2}{\pi} \int_0^\infty dx \left[ \xi \cos (\xi x) - \frac{\sin(\xi x)}{x}
\right] e^{-2 x^2} W_{\rm NLO}^F (\bar{t} = x^2).  \end{equation} Because we
compute $W_{\rm NLO}^F (\bar{t})$ numerically from the finite parameter
integrals, also ${\cal F}_{\rm NLO}^F (r;t)$ is only known numerically for a
given value of $\xi = r/\sqrt{t}$.  We show ${\cal F}_{\rm NLO}^F (r;t)$ as a
function of $\xi$ in figure~\ref{fig:FNLOplot}. 
A function that approximates ${\cal F}_{\rm
NLO}^F (r;t)$ better than 1\% in the region  $1 < \xi < 10$ and better than
2\% in the region  $\xi> 10$ is \begin{equation} \label{eq:fnloapprox}
\sum_{n=1}^{10} \frac{c_n}{n!} \left(\frac{\xi}{1+\xi/C_a} \right)^n e^{-\xi} +
\frac{44-C_b}{C_a+\xi^2} + \frac{C_b \xi^2}{\xi^4 + C_c}, \end{equation}
 with $C_a =  109.358$, $C_b = 43.8438$, $C_c = 404.790$, and
$c_n$  listed in table~\ref{tab:c12}.  
The function \eqref{eq:fnloapprox} does not reproduce  ${\cal F}_{\rm
NLO}^F(r;t)$ for $\xi < 1$.

\begin{table}[tbp] \centering \begin{tabular}{|c|c|} \hline $c_1$ &
$-0.0501648$ \\ $c_2$ & $0.526758$ \\ $c_3$ & $-5.55177$ \\ $c_4$ & $45.8753$
\\ $c_5$ & $-147.8$ \\ $c_6$ & $463.906$ \\ $c_7$ & $-851.741$ \\ $c_8$ &
$884.315$ \\ $c_9$ & $-499.105$ \\ $c_{10}$ & $121.773$ \\ \hline \end{tabular}
\caption{\label{tab:c12} Numerical values of the coefficients $c_{n}$ appearing
in eq.~\eqref{eq:fnloapprox} that make that function reproduce ${\cal F}_{\rm
NLO}^F (r;t)$ for $1 < r/\sqrt{t} < 10$ better than 1\%, and for $r/\sqrt{t} >
10$ better than 2\%.  } \end{table}

For small values of $\xi$, we may expand the terms in the square brackets in
eq.~\eqref{eq:fnloalt} as series in $\xi$ and compute the coefficients
numerically to find \begin{equation} \label{eq:fnloapprox2} {\cal F}_{\rm
NLO}^F (r;t) = 0.304930 \, \xi^3 -0.0332202 \, \xi^5 + 0.00181358 \, \xi^7 +
O(\xi^9).  \end{equation} The truncated series agrees with the numerical
calculation of ${\cal F}_{\rm NLO}^F (r;t)$ for $\xi < 0.5$ by better than
$10^{-4}$, and for $0.5 < \xi < 1.0$ by better than $5 \times 10^{-4}$. 

For large values of $\xi$ (small $t$, fixed $r$), we derive the asymptotic
behavior of ${\cal F}_{\rm NLO}^F (r;t)$ based on the behavior of $W_{\rm
NLO}^F(\bar{t})$ at small $\bar{t}$ and the Riemann--Lebesgue lemma.  We first
rewrite ${\cal F}_{\rm NLO}^F (r;t)$ as \begin{equation} \label{eq:fnloalt2}
{\cal F}_{\rm NLO}^F (r;t) = - \frac{2}{\pi} \int_0^\infty dx \left[ \xi \cos
(\xi x) f(x) - \sin(\xi x) g(x) \right], \end{equation} where $f(x) = e^{-2
x^2} W_{\rm NLO}^F (\bar{t}=x^2)$, and $g(x) = f(x)/x$.  We note that $f(x)
\sim x^2 \log (x/C)$ and $g(x) \sim x \log (x/C)$ for small $x$, where $C$ is a
constant, while they both vanish exponentially at large $x$.  We first consider
the cosine term. By integrating by parts, we obtain \begin{equation} \xi
\int_0^\infty dx \, \cos (\xi x) f(x) = - \int_0^\infty dx \sin (\xi x) f'(x) =
- \frac{1}{\xi} \int_0^\infty dx \cos (\xi x) f''(x).  \end{equation} In the
first and second equalities, we used the fact that both $f(x)$ and $f'(x)$
vanish at $x=0$ and $x=\infty$, so that the boundary terms vanish.  From the
Riemann--Lebesgue lemma, we see that the integral $\displaystyle \int_0^\infty
dx \cos (\xi x) f''(x)$ vanishes in the limit $\xi \to \infty$, because
$\displaystyle \int_0^\infty dx | f''(x)|$ is finite. We use integration by
parts again to obtain \begin{equation} \xi \int_0^\infty dx \, \cos (\xi x)
f(x) = - \frac{1}{\xi^2} \sin (\xi x) f''(x) \Big|_{x=0}^{\infty} +
\frac{1}{\xi^2} \int_0^\infty dx \, \sin (\xi x) f'''(x).  \end{equation} Since
$f''(x)$ diverges logarithmically and $\sin (\xi x)$ vanishes linearly at
$x=0$, the boundary terms vanish.  The Riemann--Lebesgue lemma does not apply
to the last term, because $f'''(x)$ diverges like $1/x$.  However, if we split
the region of integration as \begin{equation} \xi \int_0^\infty dx \, \cos (\xi
x) f(x) = \frac{1}{\xi^2} \left[ \int_0^a dx \, \sin (\xi x) f'''(x) +
\int_a^\infty dx \, \sin (\xi x) f'''(x) \right], \end{equation} where $a$ is a
positive constant, the second term in the square brackets vanishes in the limit
$\xi \to \infty$, because $\displaystyle \int_a^\infty dx | f'''(x)|$ is
finite.  In the limit $\xi \to \infty$, we can see that the first term in the
square brackets converges to a finite value by approximating $W_{\rm NLO}^F
(\bar{t})$ by eq.~\eqref{eq:WNLOF_exp}.  We obtain from the Dirichlet integral
\begin{equation} \lim_{\xi \to \infty} \int_0^a dx \, \sin (\xi x) f'''(x)
\approx 2 \pi c_L, \end{equation} where $c_L = -22/3$ is the coefficient of
$\bar{t} \log \bar{t}$ in eq.~\eqref{eq:WNLOF_exp}.  Similarly, by using
integration by parts we can rewrite the sine term of eq.~\eqref{eq:fnloalt2} as
\begin{equation} - \int_0^\infty dx \, \sin (\xi x) g(x) = \frac{1}{\xi^2}
\left[ \int_0^a dx \, \sin (\xi x) g''(x)  + \int_a^\infty dx \, \sin (\xi x)
g''(x) \right], \end{equation} where the second term in the square brackets
vanishes in the limit $\xi \to \infty$, and \begin{equation} \lim_{\xi \to
\infty} \int_0^a dx \, \sin (\xi x) g''(x) \approx \pi c_L.  \end{equation}
From these results we obtain the following asymptotic behavior for large
$\xi$ given by \begin{equation} {\cal F}_{\rm NLO}^F (r;t) \approx - \frac{6
c_L}{\xi^2} .  \end{equation} This shows that ${\cal F}_{\rm NLO}^F (r;t)$ for
a given $r$ vanishes in the limit $t \to 0$ linearly as $-6 c_L t/r^2$.  We
recall that the $t \to 0$ limit is the relevant limit, the one that reduces the
gradient flow to QCD.

We now show the numerical results for the static force in gradient flow, $r^2
F(r;t)$, at NLO in $\alpha_s$.  We compute $\alpha_s$ in the $\overline{\rm
MS}$ scheme at the scale $\mu = (r^2+8 t)^{-1/2}$ by using {\sf
RunDec}~\cite{Chetyrkin:2000yt} at four loops, and set $n_f = 4$.  At tree
level, $r^2 F(r; t=0)$ equals $\alpha_s C_F$, so that it increases with $r$,
due to the evolution of $\alpha_s (\mu=1/r)$.

\begin{figure}[ht] \begin{center}
\includegraphics[width=0.48\textwidth]{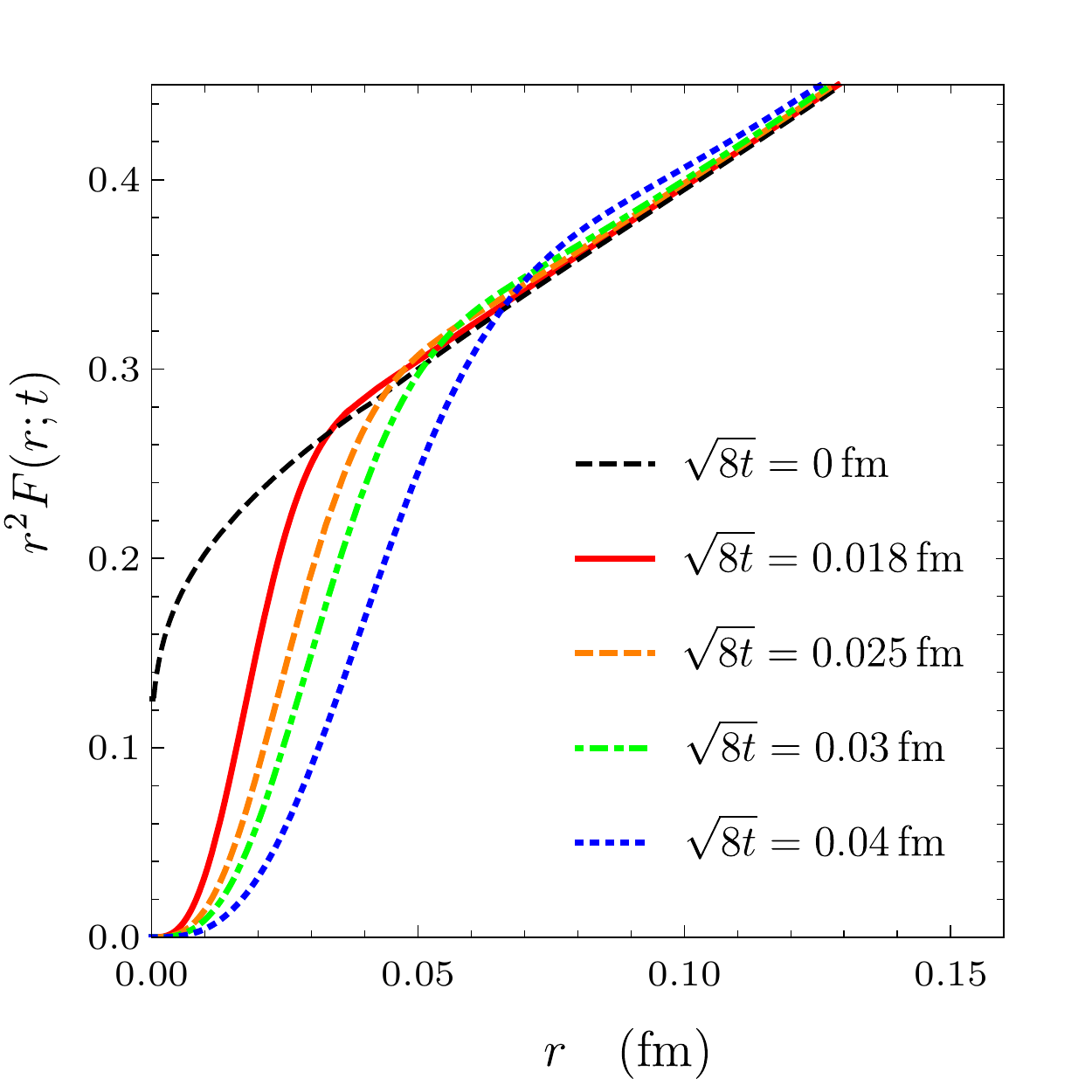}
\includegraphics[width=0.48\textwidth]{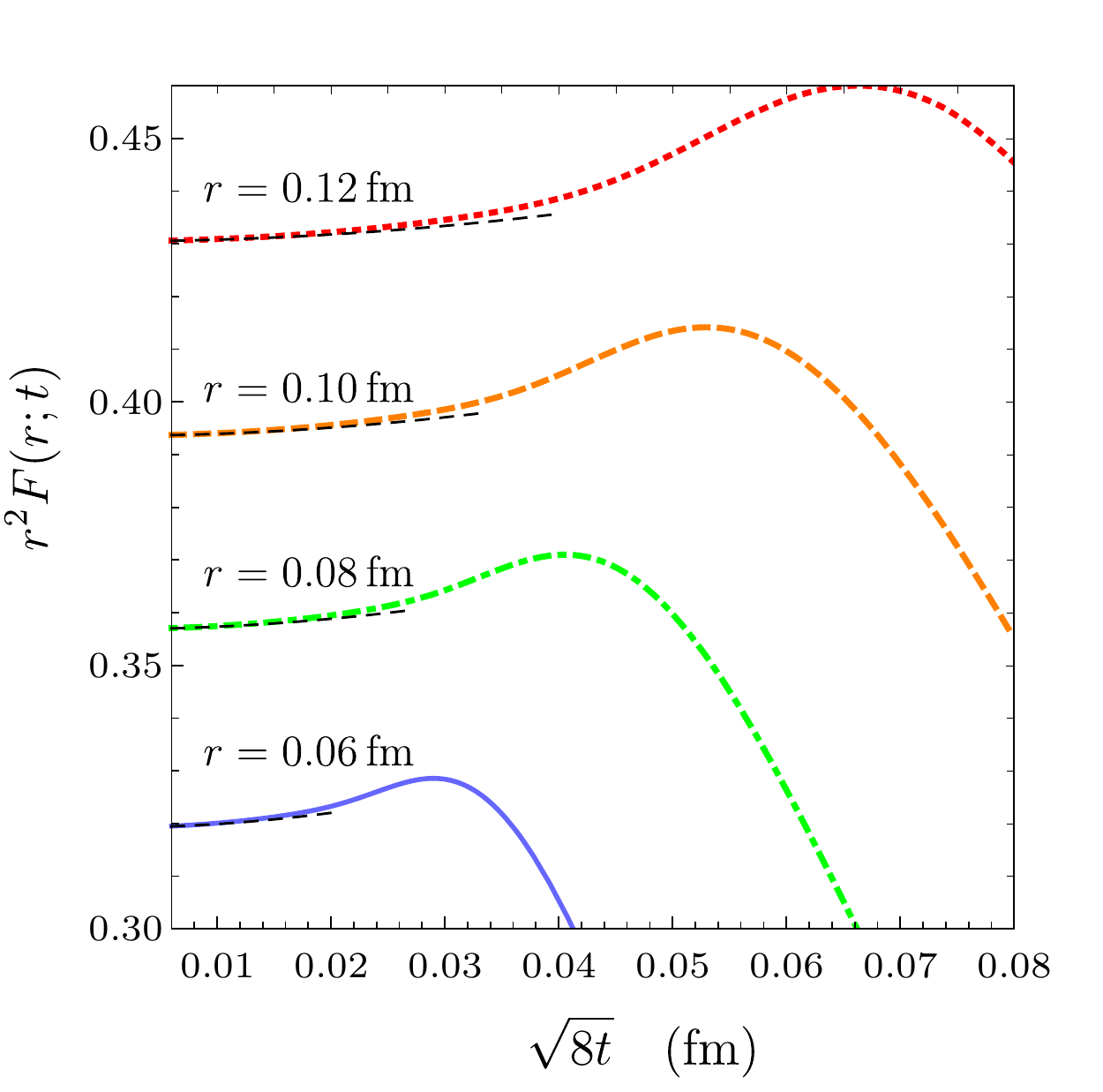}
\caption{\label{fig:rtdepplot} Left panel: numerical results for $r^2 F(r;t)$
for fixed values of $\sqrt{8 t}$ as functions of $r$.  Right panel: numerical
results for $r^2 F(r;t)$ for fixed values of $r$ as functions of $\sqrt{8 t}$;
the black dashed lines are approximate results based on eq.~\eqref{eq:asym},
which is valid at small flow time.  We have set $\mu = (r^2+8 t)^{-1/2}$ and
$n_f = 4$.  } \end{center} \end{figure}

In the left panel of figure~\ref{fig:rtdepplot}, we show $r^2 F(r;t)$ as a
function of $r$ for several fixed values of $\sqrt{8 t}$.  We can see that $r^2
F(r;t)$ vanishes for $r \to 0$, while we recover the QCD result $r^2 F(r;t=0)$
(black dashed line) for $r \gg \sqrt{8 t}$.  We observe that $r^2 F(r;t)$
slightly overshoots the QCD result before converging to the $t \to 0$ limit,
which comes from the positive correction due to ${\cal F}_{\rm NLO}^F (r;t)$
being larger than the negative correction due to ${\cal F}_{\rm
NLO}^L(r;t;\mu)$. 

In the right panel of figure~\ref{fig:rtdepplot}, we show $r^2 F(r;t)$ as a
function of $t$ for several fixed values of $r$.  For large $t$ ($\sqrt{8
t}\gtrsim r$), $r^2 F(r;t)$ is below the QCD result (given by the $t \to 0$
limit).  As $t$ decreases, $r^2 F(r;t)$ slightly overshoots the QCD result
before converging to it.  Again, this overshoot comes mainly from the positive
correction due to ${\cal F}_{\rm NLO}^F$.

By using the asymptotic expansions of ${\cal F}_{\rm NLO}^L (r;t;\mu)$ and
${\cal F}_{\rm NLO}^F (r;t)$ for small $t$, we can derive the behavior of $r^2
F(r;t)$ near $t=0$ as \begin{equation} \label{eq:asym} r^2 F(r; t) \approx r^2
F(r;t=0) + \frac{\alpha_s^2 C_F}{4 \pi} \left[ - 12 \beta_0 - 6 C_A c_L\right]
\frac{t}{r^2}, \end{equation} where $c_L = -22/3$ and $F(r;t=0)$ is the usual
QCD result for the static force \begin{equation} F(r; t=0) =
\frac{\alpha_s(\mu) C_F}{r^2} \left\{ 1+ \frac{\alpha_s}{4 \pi} \left[ a_1 +2
\beta_0 \log (\mu r e^{\gamma_{\rm E}-1}) \right]\right\} + O(\alpha_s^3).
\end{equation} For any positive $n_f$, the coefficient of the $t/r^2$ term is
positive, and grows linearly with increasing $n_f$: specifically it is $
\left[- 12 \beta_0 - 6 C_A c_L \right] = 8 n_f$. Surprisingly the coefficient
of the $t/r^2$ term vanishes at NLO in the pure SU(3) gauge theory ($n_f=0$).
We compare the exact NLO result for $r^2 F(r;t)$ at small $t$ with the
expression given in eq.~\eqref{eq:asym}, in the right panel of
figure~\ref{fig:rtdepplot}.  The function \eqref{eq:asym}, which is quadratic
in $\sqrt{8 t}$ for $n_f=4$, is represented by the dashed black curve.  The
exact result approaches smoothly the approximate one at small flow time.

\begin{figure}[ht] \begin{center}
\includegraphics[width=0.48\textwidth]{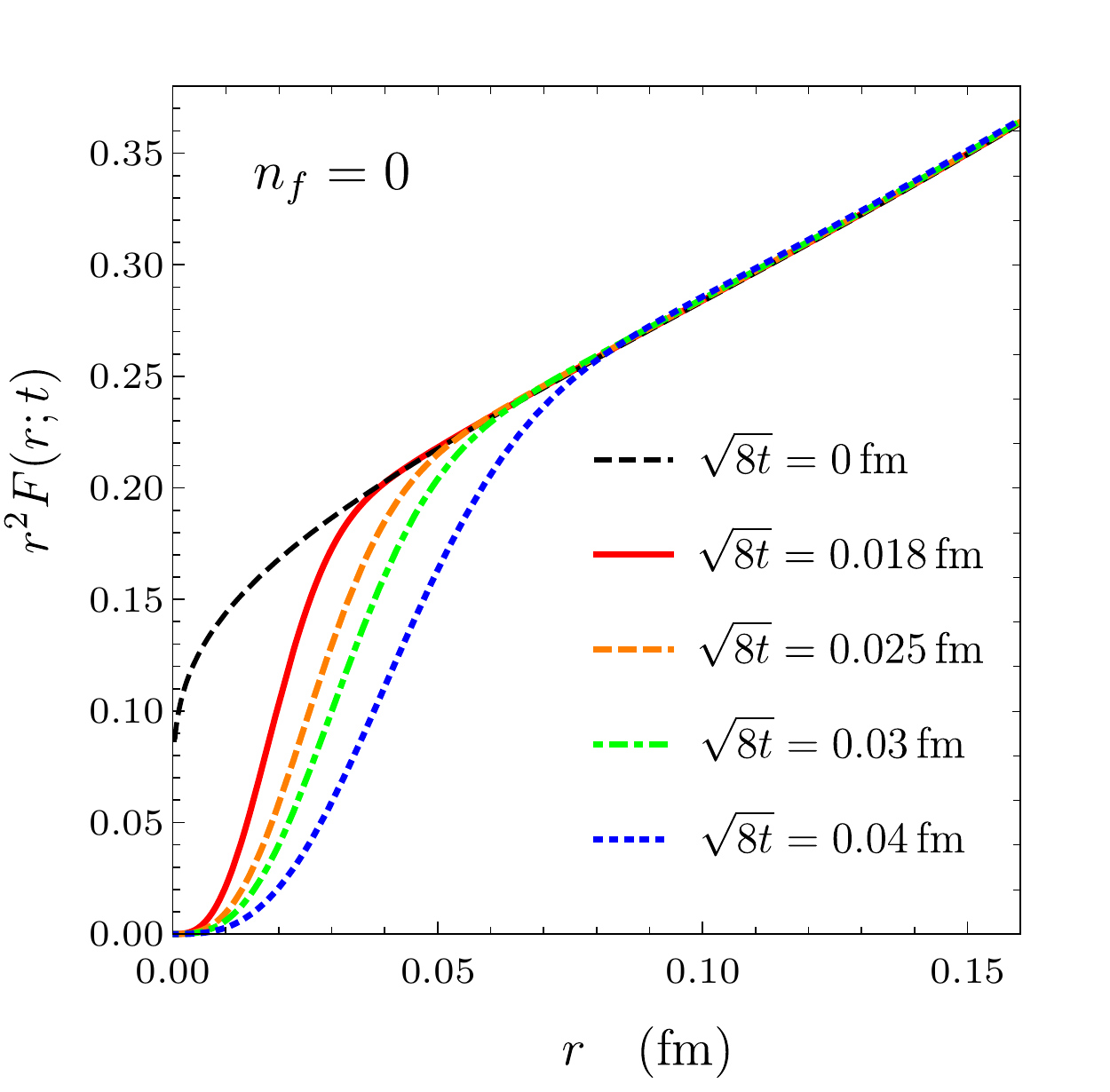}
\includegraphics[width=0.48\textwidth]{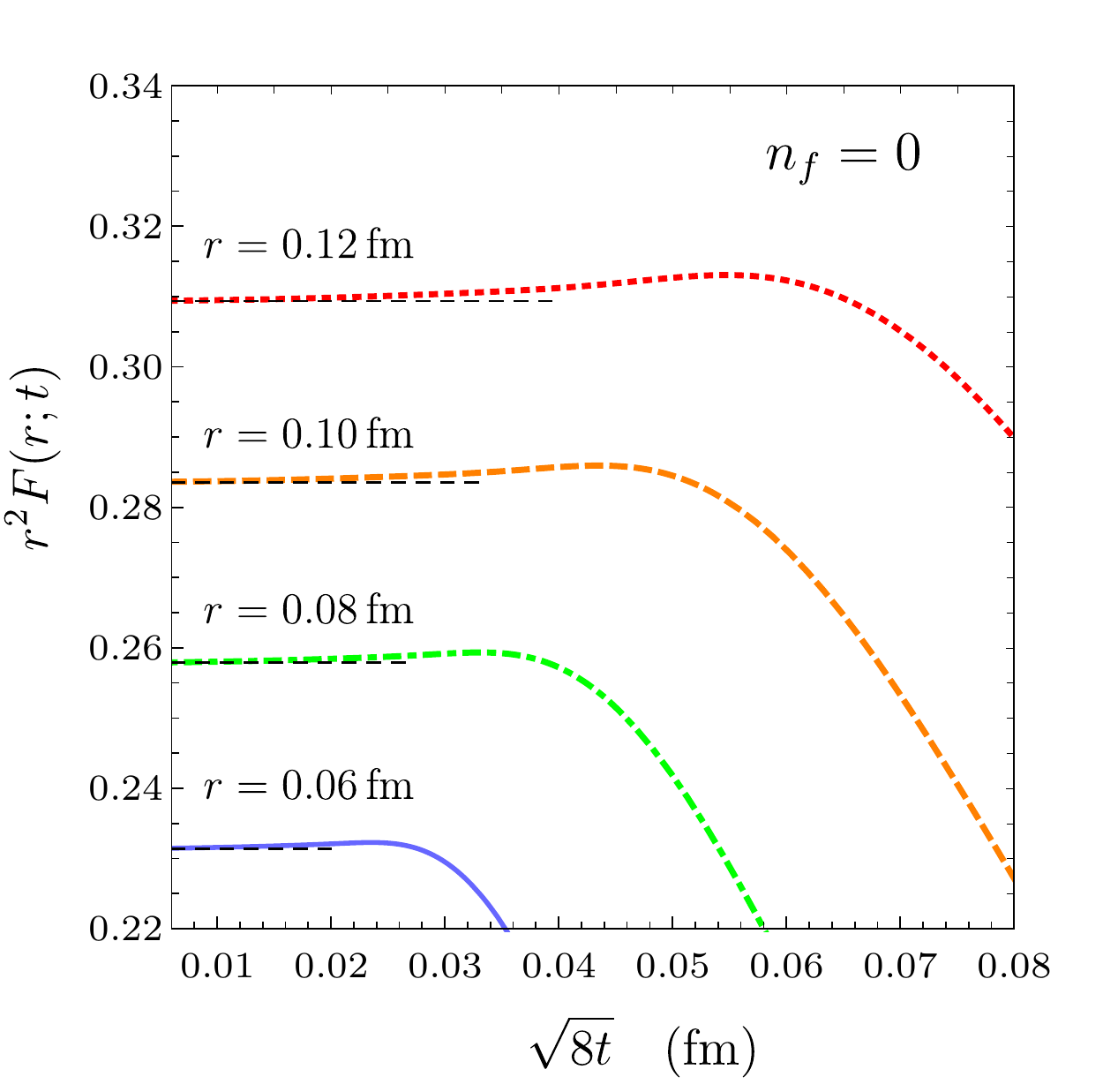}
\caption{\label{fig:rtdepplot_quenched} Left panel: numerical results for $r^2
F(r;t)$ in the pure SU(3) gauge theory ($n_f =0$) for fixed values of $\sqrt{8
t}$ as functions of $r$.  Right panel: numerical results for $r^2 F(r;t)$ in
the pure SU(3) gauge theory ($n_f =0$) for fixed values of $r$ as    functions
of $\sqrt{8 t}$; the black dashed lines are the QCD results for $r^2 F(r;t=0)$.
We have set $\mu = (r^2+8 t)^{-1/2}$.  } \end{center}
\end{figure}

We also show $r^2 F(r;t)$ in the pure SU(3) gauge theory ($n_f=0$) in
figure~\ref{fig:rtdepplot_quenched} as a function of $r$ for fixed values of
$\sqrt{8t}$, and as a function of $\sqrt{8 t}$ for fixed values of $r$.  We
have computed $\alpha_s$ in the pure SU(3) gauge theory by using {\sf
RunDec}~\cite{Chetyrkin:2000yt} at four loops, based on the value $r_0
\Lambda_{\rm QCD} = 0.637^{+0.032}_{-0.030}$ in ref.~\cite{Brambilla:2010pp},
with $r_0 = 0.5$~fm~\cite{Necco:2001xg}.  Due to the vanishing of the term
linear in $t$ in eq.~\eqref{eq:asym}, the expression in eq.~\eqref{eq:asym} is
equal to the QCD result $r^2 F(r;t=0)$, which we show in the right panel of
figure~\ref{fig:rtdepplot_quenched} as horizontal black dashed lines.

\section{\boldmath Summary and discussion} \label{sec:summary} 

In this work, we have computed the QCD static force and potential in
gradient flow to next-to-leading order (NLO) accuracy in the strong coupling.
The details of the NLO calculation have been  given in
section~\ref{sec:nlo}.  The results in momentum space have been summarized in
section~\ref{res:mom}, and the position-space results for the force have
been given in section~\ref{res:pos}.  As we have anticipated in the
introduction, the gradient flow makes the Fourier transform of the static force
in momentum space better converging, because the nonzero flow time $t$
introduces an exponentially decreasing factor in the integrand.  Thus, we
expect that the use of gradient flow may as well improve the convergence
towards the continuum limit of the lattice QCD calculation of the static force
done at finite flow time.  This is supported by the preliminary results in
ref.~\cite{Leino:2021-157/21}. 

Once the continuum limit of the lattice calculation of the force at finite flow
time has been reached, the position-space results presented in
section~\ref{res:pos} may be useful for extrapolating to zero flow time, i.e.
to QCD.  We have shown explicitly that, indeed, both momentum and
position-space NLO results reduce to the usual QCD result in the $t \to 0$
limit.  We have also derived the NLO behavior of the static force in gradient
flow near $t=0$, which is shown in eq.~\eqref{eq:asym}.  The linear correction
in $t$ vanishes in the pure SU(3) gauge theory, while it is positive for any
positive number of massless quark flavors.

The NLO correction to the static force involves the length scale $\sqrt{8 t}$,
which often appears in loop-level calculations in gradient flow.  Unlike the
case of local operator matrix elements, an optimal choice of the
renormalization scale $\mu$ is given by a combination of $1/r$ and $1/\sqrt{8
t}$, rather than just the scale $1/\sqrt{8 t}$.  This happens because the
static force depends only on the scale $1/r$ at $t=0$. 

The fact that the static force in gradient flow involves two scales $1/r$ and
$1/\sqrt{8 t}$ brings in some complications in the investigation of its
behavior near $t=0$.  While the $t \to 0$ limit exists, since the static force
is finite without renormalization of the gluon field, the NLO correction is not
analytic at $t=0$, so that we cannot expand in powers of $t$ in a
straightforward manner.  If there was only one scale $1/\sqrt{8 t}$, the
nonanalyticity in $t$ could be read off from the $\log \mu$ terms in the NLO
correction, which, in turn, are determined by the UV divergences.  However,
because of the two scales, the NLO correction involves complicated functions of
the dimensionless ratio of the two scales, which happen to be nonanalytic at
$t=0$.  As a result, the derivation of the behavior of the static force near
$t=0$ and beyond the $t \to 0$ limit in eq.~\eqref{eq:asym} required a careful
examination of the asymptotic expansion of the functions appearing in the NLO
correction. 

We expect this work to be of help in lattice QCD studies of the static force:
if gradient flow provides a better convergence of direct lattice calculations
of the force towards the continuum limit, the perturbative calculation done
here adds a better controlled zero flow time limit.  Similar analyses could be
extended to a vast range of nonperturbative quarkonium observables in the
 factorization framework provided by nonrelativistic effective field
theories~\cite{Brambilla:2004jw}.  For instance, one could study in gradient
flow the quarkonium potential at higher orders in
$1/m$~\cite{Eichten:1979pu,Barchielli:1988zp,Bali:1997am,Brambilla:2000gk,Pineda:2000sz,Brambilla:2003mu,Koma:2006si,Koma:2006fw,Koma:2010zza},
where $m$ is the heavy quark mass, static hybrid
potentials~\cite{Juge:2002br,Capitani:2018rox,Schlosser:2021wnr}, hybrid
potentials at higher orders in
$1/m$~\cite{Oncala:2017hop,Brambilla:2018pyn,Brambilla:2019jfi}, gluonic
correlators entering the expressions of quarkonium inclusive widths and cross
sections~\cite{Brambilla:2001xy,Brambilla:2002nu,Brambilla:2020xod,Brambilla:2020ojz,Brambilla:2021abf}.

\acknowledgments The work of N.~B. is supported by the DFG (Deutsche
Forschungsgemeinschaft, German Research Foundation) Grant No. BR 4058/2-2.
N.~B., H.~S.~C. and A.~V. acknowledge support from the DFG cluster of
excellence ``ORIGINS'' under Germany's Excellence Strategy - EXC-2094 -
390783311.  The work of A.~V. is funded by EU Horizon 2020 research and
innovation programme, STRONG-2020 project, under grant agreement No. 824093.
The work of X.-P.~W. is funded by the DFG Project-ID 196253076 TRR 110.

\appendix 

\section{Finite parameter integrals at one loop} \label{sec:oneloopfintable}

In this appendix, we show the expressions for the finite parameter integrals
$W_2^F(\bar{t})$, $W_4^F(\bar{t})$, $W_6^F(\bar{t})$, $W_7^F(\bar{t})$,
$W_8^F(\bar{t})$, and $W_9^F(\bar{t})$ that appear in the finite parts of the
one-loop corrections to the static potential in gradient flow, see
section~\ref{sec:nlo}.  We express them as integrals over unit hypercubes of
dimensions of up to 4.  We also show the parameter integrals for
$W_2^F(\bar{t})$, $W_8^F(\bar{t})$, and $W_9^F (\bar{t})$, even though we have
computed them analytically in terms of exponential integrals and error
functions, because the parameter integrals may be useful in some numerical
analyses.  We write $W_4^F(\bar{t})$, $W_6^F(\bar{t})$ and
$W_9^F(\bar{t})$ as sums of analytic functions of $\bar{t}$ and parameter
integrals that vanish at $\bar{t} = 0$, so that the $\bar{t} \to 0$ limit can
be easily identified.  The expressions read: \begin{eqnarray} W_2^F(\bar{t})
&=& - 2 \int_0^1 dx_1 dx_2 \Bigg[ \frac{ \bar{t} x_2 \exp \left(\frac{\left(3
x_1 x_2^2-4 \left(x_1+1\right) x_2+4\right) \bar{t}}{2 x_1 x_2^2-3
\left(x_1+1\right) x_2+4}\right)}{(x_2-1) \left(x_1 x_2-1\right) \left(2 x_1
x_2^2-3 \left(x_1+1\right) x_2+4\right)} \nonumber \\ && \hspace{20ex}
+\frac{\bar{t} e^{\frac{x_1 \bar{t}}{x_1-1}}}{(x_2-1) \left(x_1-1\right){}^2} -
\log \bar{t} - \gamma_{\rm E} \Bigg], \end{eqnarray}

\begin{eqnarray} W_4^F(\bar{t}) &=& 3 \log (2 \bar{t})+ 3 \gamma_{\rm E} +
\frac{5}{2} \nonumber \\ && + \frac{1}{2} \int_0^1 dx_1 dx_2 dx_3 \Bigg[
\frac{8 \exp \left(\frac{x_2 \left(x_1 x_2-1\right) \bar{t}}{x_1 \left(2
x_2-1\right)-2}\right)}{x_2 \left(-2 x_2 x_1+x_1+2\right){}^2}+ \frac{8
e^{\frac{x_1^2 \left(x_2-1\right) x_2 \bar{t}}{2 x_1
\left(x_2-1\right)-1}}}{x_2 \left(1-2 x_1 \left(x_2-1\right)\right){}^2}
\nonumber\\ && + \frac{16 x_2 \left(x_3-1\right) \left(x_2 x_3-1\right) \exp
\left(\frac{x_2 x_3 \left(x_1^2 x_2 \left(x_3-1\right) \left(x_2
x_3-1\right)-1\right) \bar{t}}{2 x_1 \left(x_3-1\right) x_3 x_2^2-2 x_1
\left(x_3-1\right) x_2-2 x_3 x_2+x_2+1}\right)}{x_3 \left(2 x_1
\left(x_3-1\right) x_3 x_2^2-2 x_1 \left(x_3-1\right) x_2-2 x_3
x_2+x_2+1\right){}^3} \nonumber \\ && +\frac{16 x_2 \left(x_1 x_2 x_3-1\right)
\left(x_2 x_3-1\right) \exp \left(\frac{x_3 \left(-x_2 x_3+x_1 x_2 \left(x_2
\left(x_3^2-1\right)-x_3\right)+1\right) \bar{t}}{2 x_1 \left(x_3-1\right) x_3
x_2^2-\left(x_1+1\right) \left(2 x_3-1\right) x_2+2}\right)}{x_3 \left(2 x_1
\left(x_3-1\right) x_3 x_2^2-\left(x_1+1\right) \left(2 x_3-1\right)
x_2+2\right){}^3} \nonumber \\ && + \frac{16 x_2 \left(x_3-1\right) \left(x_1
x_2 x_3-1\right) \exp \left(\frac{x_2 x_3 \left(-x_3 x_2+x_2+x_1 \left(x_2^2
\left(x_3-1\right) x_3-1\right)\right) \bar{t}}{2 x_1 \left(x_3-1\right) x_3
x_2^2+\left(-2 x_3 x_1+x_1-2 x_3+2\right) x_2+1}\right)}{x_3 \left(2 x_1
\left(x_3-1\right) x_3 x_2^2+\left(-2 x_3 x_1+x_1-2 x_3+2\right)
x_2+1\right){}^3} \nonumber \\ && -\frac{16 x_2}{x_3 \left(x_1
x_2+x_2+2\right){}^3}-\frac{16x_2 }{x_3\left(2 x_1
x_2+x_2+1\right){}^3}-\frac{16 x_2}{x_3 \left(\left(x_1+2\right)
x_2+1\right){}^3} \nonumber\\ && +\frac{4 \bar{t} \exp
\left(\frac{\left(\left(x_1-1\right) \left(x_2-1\right) x_3^2-x_1 x_2\right)
\bar{t}}{-2 x_2 x_1+x_1+x_2+2 \left(x_1-1\right) \left(x_2-1\right)
x_3}\right)}{\left(-2 x_2 x_1+x_1+x_2+2 \left(x_1-1\right) \left(x_2-1\right)
x_3\right){}^2} \nonumber \\ && -\frac{8}{x_2 \left(2
x_1+1\right){}^2}-\frac{8}{x_2 \left(x_1+2\right){}^2} -2 \log
\left(\frac{81}{2}\right) \Bigg], \end{eqnarray}

\begin{eqnarray} W_6^F(\bar{t}) &=& 2 \log (2\bar{t}) + 2 \gamma_{\rm E}
-\frac{1}{2}  \nonumber \\ && + \frac{1}{6} \int_0^1 dx_1 dx_2 dx_3 \Bigg[
\frac{48 \left(x_2-1\right) x_1 \exp \left(\frac{x_1^2 \left(x_2-1\right) x_2
\left(x_3-1\right){}^2 \bar{t}}{2 x_1 \left(x_2-1\right)-1}\right)}{x_2 \left(2
x_1 \left(x_2-1\right)-1\right){}^3} \nonumber \\ && + \frac{48 \left(x_1
x_2-1\right) \exp \left(\frac{x_2 \left(x_1 x_2-1\right) \left(x_3-1\right){}^2
\bar{t}}{x_1 \left(2 x_2-1\right)-2}\right)}{x_2 \left(x_1 \left(2
x_2-1\right)-2\right){}^3} \nonumber \\ && +\frac{24 x_2
\left(\left(x_1-1\right) x_2 \left(x_3+3\right)-2 x_1\right) \bar{t} \exp
\left(\frac{\left(x_1-1\right) x_2^2 \left(x_3-1\right){}^2 \bar{t}}{2
\left(x_1-1\right) x_2-x_1}\right)}{\left(2 \left(x_1-1\right)
x_2-x_1\right){}^3} \nonumber \\ && +\frac{6 \left(e^{\frac{1}{2}
\left(x_1-1\right){}^2 x_2 \bar{t}}-1\right)}{x_2} - \frac{48 x_1}{x_2 \left(2
x_1+1\right){}^3}-\frac{48}{x_2 \left(x_1+2\right){}^3} +4 -6 \log 3 \Bigg],
\end{eqnarray}

\begin{eqnarray} W_7^F(\bar{t}) &=& 4 \int_0^1 dx_1 dx_2 dx_3 dx_4  \Bigg[
\frac{\bar{t} \exp \left(\frac{\bar{t} \left(x_1-1\right)
\left(x_2+x_3\right){}^2}{-x_1+2 \left(x_1-1\right) x_2+2 \left(x_1-1\right)
x_3}\right)}{\left(x_1-2 \left(x_1-1\right) x_2-2 \left(x_1-1\right)
x_3\right){}^2} \nonumber \\ && \hspace{-1cm} + \frac{\bar{t}
\left(x_1-1\right) \left(x_2-1\right) \exp \left(\frac{\bar{t}
\left(\left(x_1-1\right) \left(x_2-1\right) x_3^2+2 \left(x_1-1\right)
\left(x_2-1\right) x_4 x_3+\left(x_1-1\right) \left(x_2-  1\right) x_4^2-x_1
x_2\right)}{-2 x_2 x_1+x_1+x_2+2 \left(x_1-1\right) \left(x_2-1\right) x_3+2
\left(x_1-1\right) \left(x_2-1\right) x_4}\right)}{2 \left(-2 x_2 x_1+x_1+x_2+2
\left(x_1-1\right) \left(x_2-1\right) x_3+2 \left(x_1-1\right)
\left(x_2-1\right) x_4\right){}^3} \Bigg], \nonumber \\ \end{eqnarray}

\begin{eqnarray} W_8^F(\bar{t}) = 2 \int_0^1 dx_1 dx_2 dx_3 \frac{\bar{t} \exp
\left(-\frac{\bar{t} \left(\left(x_1-1\right) \left(x_2-     1\right) x_3^2+2
\left(x_1-1\right) \left(x_2-1\right) x_3-x_1-x_2+1\right)}{x_1+x_2-2
\left(x_1-1\right) \left(x_2-1\right) x_3-   2}\right)}{\left(x_1+x_2-2
\left(x_1-1\right) \left(x_2-1\right) x_3-2\right){}^2}, \end{eqnarray}

\begin{eqnarray} W_9^F(\bar{t}) = -1 - 2 \int_0^1 dx_1 dx_2
\frac{e^{\frac{\left(x_1-1\right) \left(x_2-1\right){}^2
\bar{t}}{x_1-2}}-1}{\left(x_1-2\right){}^2}.  \end{eqnarray}

\bibliography{force.bib} \bibliographystyle{JHEP}

\end{document}